\documentclass[prl,superscriptaddress,floatfix,amsmath,notitlepage,twocolumn]{revtex4-1}
\usepackage{graphicx}
\usepackage{color}
\usepackage{siunitx}
\usepackage[normal]{subfigure}
\usepackage[usenames,dvipsnames]{xcolor}
\usepackage{amssymb}
\usepackage{dsfont}
\newcommand{\be}{\begin{equation}}
\newcommand{\ee}{\end{equation}}

\newcommand*\pFqskip{8mu}
\catcode`,\active
\newcommand*\pFq{\begingroup
        \catcode`\,\active
        \def ,{\mskip\pFqskip\relax}%
        \dopFq
}
\catcode`\,12
\def\dopFq#1#2#3#4#5{%
        {}_{#1}F_{#2}\biggl(\genfrac..{0pt}{}{#3}{#4};#5\biggr)%
        \endgroup
}

\newmuskip\pFqskip
\mathchardef\pFcomma=\mathcode`, % keep a copy of the comma

\newcommand*\pFtildeq[5]{%
  \begingroup
  \begingroup\lccode`~=`,
    \lowercase{\endgroup\def~}{\pFcomma\mkern\pFqskip}%
  \mathcode`,=\string"8000
  {}_{#1}\tilde{F}_{#2}\left(\genfrac..{0pt}{}{#3}{#4};#5\right)%
  \endgroup
}

\usepackage[colorlinks=true,urlcolor=blue,linkcolor=blue,citecolor=blue]{hyperref}

\begin{document}

%\title{Photon--photon interactions in waveguide QED with many emitters}
\title{Strongly correlated photon transport in waveguide QED with weakly coupled emitters}

\author{Sahand Mahmoodian}
\affiliation{Institute for Theoretical Physics, Institute for Gravitational Physics (Albert Einstein Institute), Leibniz University Hannover, Appelstra{\ss}e 2, 30167 Hannover, Germany}
\affiliation{Niels Bohr Institute, University of Copenhagen, Blegdamsvej 17, DK-2100 Copenhagen, Denmark}
\author{Mantas \v{C}epulkovskis}
\author{Sumanta Das}
\author{Peter Lodahl}
\affiliation{Center for Hybrid Quantum Networks (Hy-Q), Niels Bohr Institute, University of Copenhagen, Blegdamsvej 17, DK-2100 Copenhagen, Denmark}
\author{Klemens Hammerer}
\affiliation{Institute for Theoretical Physics, Institute for Gravitational Physics (Albert Einstein Institute), Leibniz University Hannover, Appelstra{\ss}e 2, 30167 Hannover, Germany}
\author{Anders~S.~S{\o}rensen}
\affiliation{Center for Hybrid Quantum Networks (Hy-Q), Niels Bohr Institute, University of Copenhagen, Blegdamsvej 17, DK-2100 Copenhagen, Denmark}

\date{\today}

\begin{abstract}
We show that strongly correlated photon transport can be observed in waveguides containing optically dense ensembles of emitters. Remarkably, this occurs even for weak coupling efficiencies. Specifically, we compute the photon transport properties through a chirally coupled system of $N$ two-level systems driven by a weak coherent field, where each emitter can also scatter photons out of the waveguide. The photon correlations arise due to an interplay of nonlinearity and coupling to a loss reservoir, which creates a strong effective interaction between transmitted photons. The highly correlated photon states are less susceptible to losses than uncorrelated photons and have a power-law decay with $N$. This is described using a simple universal asymptotic solution governed by a single scaling parameter which describes photon bunching and power transmission.  We show numerically that, for randomly placed emitters, these results hold even in systems without chirality. The effect can be observed in existing tapered fiber setups with trapped atoms.
\end{abstract}

\maketitle

\begin{figure}[!t]
\includegraphics[width=\columnwidth]{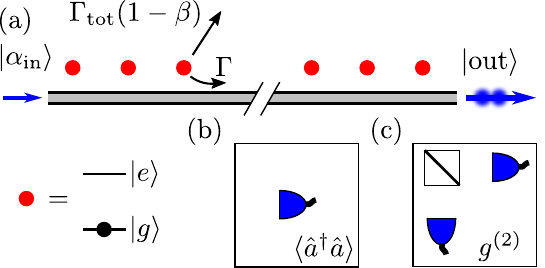}
\caption{\label{fig:schem} (a) $N$ chirally coupled two-level emitters (red circles) driven by an external coherent field $| \alpha_{\rm in} \rangle$ with a corresponding strongly correlated output photon state $|\rm{out} \rangle$. Each emitter is coupled to the waveguide with a decay rate $\Gamma = \beta \Gamma_{\rm tot}$ and to external loss modes with a decay rate $\Gamma_{\rm tot}(1-\beta)$. The output state is probed by measuring (b) power $\langle \hat{a}^\dagger \hat{a} \rangle$ or (c) the normalized second-order correlation function $g^{(2)}$.}
\end{figure}

Describing the dynamics of quantum systems that are far from equilibrium is currently one of the main challenges of physics. Considerable effort is put into understanding these systems, e.g., in quantum many-body physics and nonlinear dynamics \cite{Polkovnikov2011RMP,  Eisert2015NPHYS}, as well as developing quantum simulators to investigate them \cite{Cirac2012NPHYS}. In the field of mesoscopic physics such dynamics are especially studied through quantum transport \cite{Datta1997BOOK, DasSarma2011RMP, Laird2015RMP}. Recently, quantum transport of photons has emerged as an analogous system to study non-equilibrium  quantum dynamics in optical systems \cite{Chang2007NPHYS, Carusotto2013RMP, Chang2014NPHOT}. Most notably this has been investigated in weakly driven strongly interacting Rydberg gasses, where effective photon--photon interactions at the few-photon level have been observed \cite{Dudin2012Science, Parigi2012PRL, Peyronel2012Nature, Maxwell2013PRL, Baur2014PRL, Thompson2017Nature}.
This has led to the demonstration of fascinating new phenomena such as correlated two- \cite{Maghrebi2015PRL, Firstenberg2013Nature} and three-photon \cite{Gullans2016PRL, Jachymski2016PRL, Liang2017Science} bound states. Similar photon--photon interactions are also investigated for quantum emitters strongly coupled to optical waveguides or cavities. Here the intrinsic nonlinearity of a single emitter plays the role of a nonlinear medium \cite{Goban2014NCOM, Tiecke2014Nature, Hacker2016Nature}. Significant effort has therefore been put into creating light--matter interfaces between an emitter and a single optical mode with near-unity coupling efficiency $\beta \sim 1$ so that dissipation is minimized \cite{Arcari2014PRL, Lodahl2015RMP}. Contrary to this, we consider quantum transport through a strongly dissipative system consisting of  $N \gg 1$ quantum emitters coupled to a waveguide. We analytically compute the dynamics of this system in the case of chiral coupling \cite{Petersen2014Science, Mitsch2014NCOM, Sollner2015NNANO, Young2015PRL, Coles2016NCOM, Lodahl2017Nature, Stannigel2012NJP, Ramos2014quantum, Pichler2015PRA, Mahmoodian2016PRL, Scheucher2016Science}, where emitters only couple to photons propagating to the right (Fig.~\ref{fig:schem}(a)). Surprisingly, we find that the interplay of weak nonlinearity and strong dissipation leads to the emergence of highly nonlinear transmission and strongly correlated photon states at the output. 
Previously, such dissipatively induced photon correlations have been studied for strong optical nonlinearities \cite{Peyronel2012Nature, Zeuthen2017PRL, Murray2018PRL}.
Since these dynamics can occur even for weakly coupled emitters $\beta \ll 1$, they are readily observable in a larger range of systems, e.g., in experiments on atoms coupled to nanofibers \cite{Mitsch2014NCOM, SayrinPRX2015, Sorensen2016PRL, Corzo2016PRL, Solano2017}.

The phenomena we investigate is based on dissipation 
and can thus only be observed in optically dense ensembles. Specifically, when two resonant photons interact with the same atom, they can exchange energy, creating correlated red- and blue-detuned photons (sidebands). Since losses are strongest on resonance, resonant uncorrelated photons suffer strong loss (exponential scaling with $N$), while off-resonant correlated photons incur reduced loss.  The smallest detunings are lost first, so that the decay rate constantly decreases with subsequent atoms since the remaining photons have larger detunings. This leads to a power-law decay of the transmission and the output being dominated by strongly correlated (bunched) photons. Such power-law decay is ubiquitous in critical \cite{StanleyBOOK} or chaotic \cite{ThompsonBOOK} systems and is linked to scale invariance and the absence of a characteristic length scale, so that the microscopic details of the system becomes irrelevant. Analogously, we find that in the limit of large optical depth, the dynamics attains a universal scaling relation, which becomes independent of the precise value of the coupling efficiency. In the following we derive these results analytically assuming chiral interactions, but we show numerically that these conclusions are robust and also apply to bidirectional interactions, i.e. non-chiral, for weakly coupled randomly placed emitters.

We consider a system of $N$-chirally coupled two-level emitters
continuously driven by a weak coherent field while dissipatively
coupled to a loss reservoir (Fig.~\ref{fig:schem}(a)). The emitters are coupled to the waveguide with a decay rate $\Gamma = \beta \Gamma_{\rm tot}$ and radiate to the reservoir with the decay rate $\Gamma_{\rm tot} (1- \beta)$, where $\Gamma_{\rm tot}$ is the total decay rate of the emitters. Solving for the dynamics of this system can be approached in a variety of ways: it constitutes a cascaded quantum system \cite{Carmichael1993PRL, Gardiner1993PRL} for which a master equation can be derived \cite{Gardiner1993PRL, Pichler2015PRA,  Lodahl2017Nature}, but it is challenging to obtain general solutions as the number of emitters increases. Other approaches use a Green function to treat photon propagation, but generally require numerical solutions \cite{Asenjo2017PRA, Manzoni2017NCOM}. Here we develop an approach based on scattering matrices. We assume that the emitters are driven at a level well below saturation such that the dynamics of the system can be described by the one- and two-photon Fock states, and we thus compute the $N$-emitter scattering matrix for these manifolds. Computing the single-photon transmission is straightforward \cite{Shen2005OL}. Significant research has been put in developing two-photon scattering matrices for a single emitter \cite{Kojima2003PRA, Hofmann2003PRA, Shen2007PRA, Zheng2010PRA, Fan2010PRA, Shi2015PRA}, and generalizations to $N$-emitters have also been developed in the absence of loss \cite{Rupasov1982JETP, Rupasov1984JETP, Yudson1985JETP, Pletyukhov2012NJP, Ringel2014NJP}. Here we compute the $N$-emitter two-photon scattering matrix by projecting the input two-photon state on the scattering eigenstates, which can be determined using the Bethe ansatz technique as described in Ref.~\cite{Shen2007PRA}. Computing the $N$-emitter scattering matrix then simply requires raising the eigenvalues to the $N$-th power.

The single frequency input coherent state is expressed up to the two-photon state as $| \alpha_{\rm in} \rangle = e^{-\frac{|\alpha|^2}{2}}\left[1  + \alpha \, \hat{a}^\dagger_{k_0} +  \frac{\alpha^2}{2} \hat{a}^\dagger_{k_0} \hat{a}^\dagger_{k_0} \right] | 0 \rangle$. We linearize and rescale the waveguide dispersion and set the group velocity $v_g=1$,  such that wavenumber and frequency, as well as distance and time, have the same units. Resonant photons correspond to $k_0=0$ and $\hat{a}^\dagger_{k_0}$ creates a photon with detuning $k_0$. Unlike bidirectional systems \cite{Chang2012NJP}, in a chiral system the propagation phase between the emitters amounts to an overall phase in the Markovian limit and does not affect the dynamics \cite{Gardiner1993PRL, Lodahl2017Nature}. The $N$-emitter scattering matrix for up to two photons is $\hat{S}^N = \left[\hat{S}_{11} + \hat{S}_{22} + \hat{S}_{12} \right]^N$. Here, $\hat{S}_{11}$ and $\hat{S}_{22}$ are the one- and two-photon scattering matrices, and $\hat{S}_{12}$ describes scattering of two input photons where one is transmitted and the other is lost. This term is required when $\beta<1$. Note that to ensure that different decays add up incoherently, $\hat{S}_{12}$ contains the state of the photons which are lost. Using the orthogonality of the one- and two-photon subspaces, the scattering matrix restricted to one and two-photons is
\begin{equation}
\label{eq:scattering}
\hat{S}^N = \hat{S}_{11}^N + \hat{S}_{22}^N + \sum_{M=0}^{N-1} \hat{S}_{11}^{N-M-1}\hat{S}_{12}\hat{S}_{22}^M,
\end{equation}
and we define contributions to the output state as $\hat{S}^N | \alpha_{\rm in} \rangle \equiv | {\rm out} \rangle_1 +| {\rm out} \rangle_2 +| {\rm out} \rangle_{21}$. Here, we only consider the part of the scattering with outgoing photons. Computing $\hat{S}_{11}^N | \alpha_{\rm in} \rangle$ is simple: since the scattering matrix must conserve the photon energy, it simply multiplies the creation operator by a transmission coefficient: $a^\dagger_{k} \rightarrow t_k^N a^\dagger_{k}$ with $t_k = 1-2\beta/(1- 2i k/\Gamma_{\rm tot})$ \cite{Fan2010PRA}. This is equivalent to scattering off a single-sided cavity. Consequently, the linear contribution to the output power scales exponentially with $N$, $\langle a^\dagger a \rangle_1 = |t_{k_0}|^{2N} |\alpha|^2/L $, where $L$ is a quantization length. Thus the linear single-photon response yields the usual exponential decay with $N$ when $|t_{k_0}|<1$.

Computing $\hat{S}_{22}^N | \alpha_{\rm in} \rangle$ is more involved. We do this by projecting the input state on the orthonormal set of two-photon scattering eigenstates computed in \cite{Shen2007PRA}. These consist of a set of extended states $| W_{E, \Delta} \rangle$, with position space representation $W_{E,\Delta} (x_c,x) = \sqrt{2} e^{i E x_c} \left[2 \Delta \cos{\Delta x - \Gamma \operatorname{sgn}{(x)}\sin{\Delta x}} \right]/(2\pi \sqrt{4 \Delta^2 + \Gamma^2})$, where for two photon positions $x_1$ and $x_2$, the centre of mass and difference coordinates are $x_c = (x_1+x_2)/2$, and $x=x_1-x_2$ \cite{Shen2007PRA}. The two indices of $W$ are a two-photon detuning $E=k+p$ and a frequency difference of the two-photons $\Delta=\frac{k - p}{2}$, where $k$ and $p$ are the detunings of the two photons. The remaining eigenstates are a set of bound states $| B_{E} \rangle$ with $B_{E}(x_c,x) = \sqrt{\frac{\Gamma}{4\pi}} e^{i E x_c}e^{- \Gamma/2 |x|}$, which only vary with the two-photon detuning $E$.

The two-photon scattering matrix operating on the input state gives \cite{Shen2007PRA}
\begin{equation}
\label{eq:twoPhotScatt}
\begin{split}
%| {\rm out} \rangle_2 =  \frac 1 2 \int dE d\Delta \, t_{\frac E 2 + \Delta}^N t_{\frac E 2 - \Delta}^N  | W_{E, \Delta} \rangle \langle W_{E, \Delta} | \alpha_{\rm in} \rangle\\
% + \int dE \, \tilde{t}_E^N | B_{E} \rangle \langle B_{E} | \alpha_{\rm in} \rangle\\
\frac{| {\rm out} \rangle_2}{A} =  \tilde{t}_{2k_0}^N c_1 |  B_{2k_0} \rangle  - \int \frac{d \Delta \, t_{k_0 + \Delta}^N t_{k_0 - \Delta}^N}{\Delta \sqrt{1+ 4 \frac{\Delta^2}{\Gamma^2}}} | W_{2k_0, \Delta} \rangle ,
\end{split}
\end{equation}
where, henceforth, integrals range over $\mathds{R}$, $c_1 = \sqrt{\frac{8\pi}{\Gamma}}$,  and $\tilde{t}_E = 1-4\beta/(1+\beta - i E/\Gamma_{\rm tot})$. Additionally, $A=\alpha^2/L \, e^{-\frac{|\alpha|^2}{2}} \sim P_{\rm in}$ where the input power is $P_{\rm in} = \alpha^2/L$, and we henceforth take $\alpha$ to be real. Using these eigenstates, we obtain a position representation of the full two-photon output state by performing the integral over $\Delta$ in (\ref{eq:twoPhotScatt}). The special case of $\beta=1$ has previously been treated and leads to a parity effect in the output state for a resonant drive \cite{Stannigel2012NJP, Ringel2014NJP}.
%This integral is computed analytically by defining a contour and using Cauchy's method of residue for the $N$-th order poles.
The full two-photon output state is
\begin{figure}[!t]
\includegraphics[width=\columnwidth]{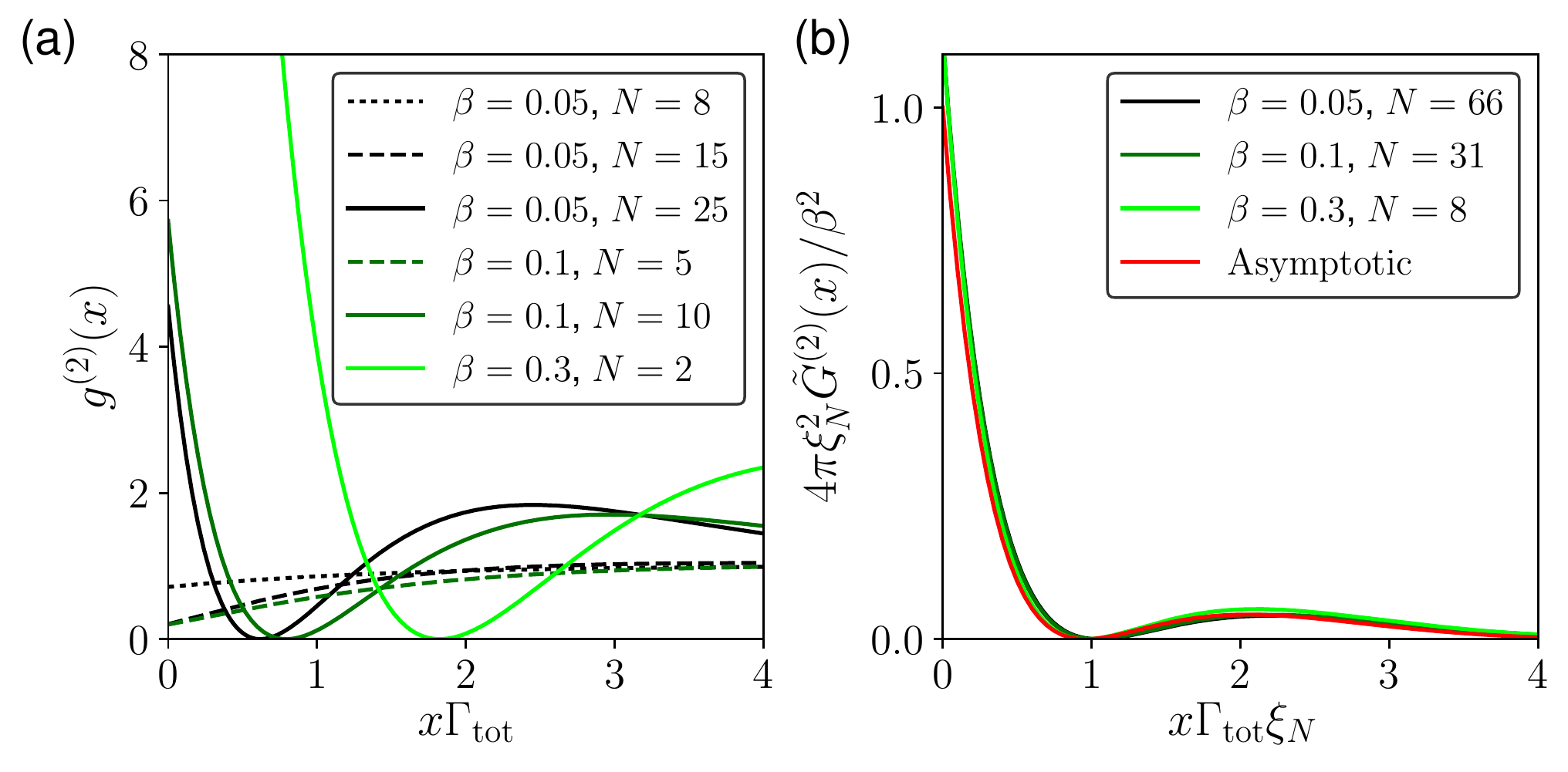}
\caption{\label{fig:g2} (a) Normalized second-order correlation function $g^{(2)}(x)$ for different numbers of emitters $N$ and coupling efficiencies $\beta$. As the optical depth increases the correlation function becomes strongly bunched. (b) Second-order correlation function $\tilde{G}^{(2)}(x)$ scaled by $4\pi \xi_N^2/\beta^2$. The emitter numbers are chosen so that the linear transmission is $(1-2\beta)^{2 N}\sim 10^{-6}$.}
\end{figure}
\begin{equation}
\label{eq:twoPhotOut}
\begin{split}
|\textrm{out}\rangle_2 &=  \frac{A}{2}\int d x_1 d x_2 \, \hat{a}^{\dagger}(x_1) \hat{a}^{\dagger}(x_2) | 0 \rangle \psi_N(x_c,x),
\end{split}
\end{equation}
with $\psi_N(x_c,x) = e^{2i k_0 x_c}\left[t_{k_0}^{2N} - \phi_N(x) \right]$. The $t_{k_0}^{2N}$ term corresponds to uncorrelated photons interacting individually with all $N$ emitters while $\phi_N(x)$ contains the photon correlations induced by the interactions. We calculate the correlations $\phi_N(x)$ analytically, but for brevity, we leave the exact form to the Supplementary Material (SM) and only show its asymptotic form below. The correlations induced by the photon--photon interactions are quantified by the normalized second-order correlations function $g^{(2)}(x) = \langle \hat{a}^\dagger(0) \hat{a}^\dagger(x) \hat{a}(x) \hat{a}(0) \rangle/ \langle \hat{a}^\dagger \hat{a} \rangle = |\psi_N(x_c,x)|^2/|t_{k_0}|^{4N} + O(\frac{P_{\rm in}}{P_{\rm sat}})$, where the saturation power is $P_{\rm sat} =\Gamma_{\rm tot}/\beta$. Throughout the remainder of this manuscript we consider a resonant drive $k_0=0$ as it generates the most interesting physics.

\begin{figure}[!t]
\includegraphics[width=\columnwidth]{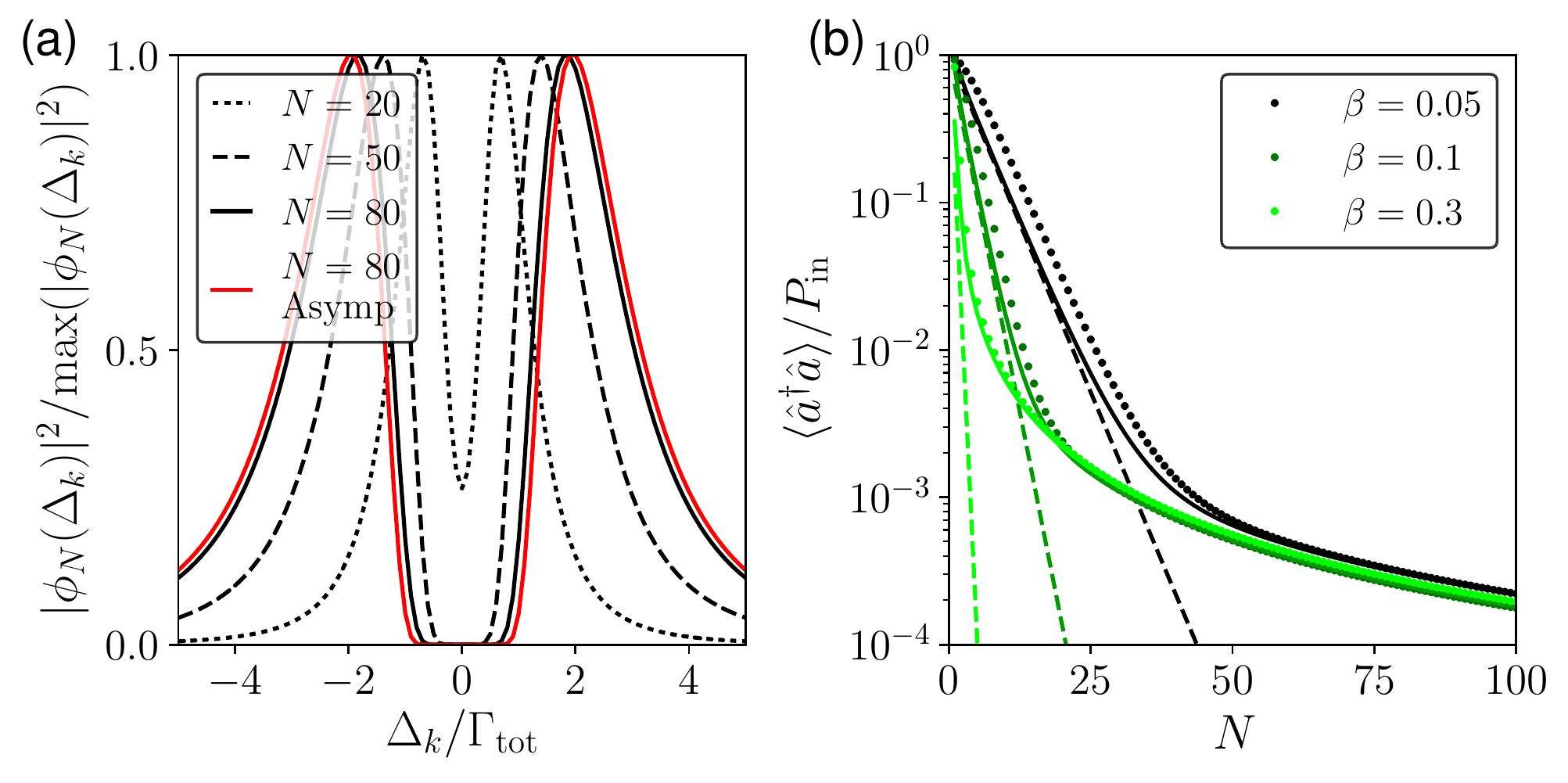}
\caption{\label{fig:power} (a) The normalized magnitude squared of the Fourier Transform of the correlated part of the two-photon wavepacket $\phi_N$ for $\beta=0.05$ with the asymptotic expression plotted for $N=80$ in red. (b) Normalized output intensity $\langle \hat{a}^\dagger \hat{a} \rangle / P_{\rm in}$ versus emitter number $N$. Broken lines show the linear output intensity $(1-2\beta)^{2 N}$ for uncorrelated photon transport while the solid lines show the asymptotic scaling. For large optical depths the transmitted power shows a power-law decay $N^{-3/2}$. The input power is $P_{\rm in}=0.1 P_{\rm sat}$.}
\end{figure}

Figure \ref{fig:g2}(a) shows $g^{(2)}(x)$ for different $\beta$ and $N$. As $N$ increases, $g^{(2)}(x)$ becomes strongly bunched even for $\beta \ll 1$. This signifies that the output contains strong photon--photon correlations and happens because the linear component of the transmitted power $\sim |t_{k_0}|^{2N}$ decays exponentially with $N$ while $\phi_N(x)$ does not. We can understand this by considering the Fourier transform of the correlated part of the two-photon wavepacket $\phi_N(\Delta_k)$, where $\Delta_k=(k_1-k_2)/2$, which we show in Fig.~\ref{fig:power}(a). Nonlinear interactions generate correlated frequency sidebands with $\Delta_k \neq 0$. Meanwhile the loss of the system is strongest on resonance and thus frequency components $\Delta_k \sim 0$  suffer strong loss. This leads to a two-lobed shape in Fourier space whose inverse Fourier transform determines the shape of $g^{(2)}(x)$. The detuning of the peaks of $\phi_N(\Delta_k)$ increase with $N$, and thus loss due to each subsequent emitter decreases and the scaling of $\phi_N$ is sub-exponential. We highlight that this occurs for all $\beta < 1$ provided the optical depth is large. The slow decay of $\phi_N$ and the resulting large values $g^{(2)}(0)$ reveal that the transmission of the system is dominated by events where two simultaneously incident photons form a correlated state. For sufficiently large optical depth, photons interacting individually will be completely blocked and the transmission is therefore dominated by two-photon events leading to strong photon bunching.

We now derive an asymptotic expression for the non-exponentially decaying parts of $\phi_N$. Since detuned Fourier components dominate, we expand the second term in (\ref{eq:twoPhotScatt}) to second order in $\Gamma_{\rm tot}/\Delta$ and get $t_{\Delta}^N t_{-\Delta}^N \sim \exp{\left[-\Gamma_{\rm tot}^2 \xi_N^2/\Delta^2 \right]}$, where $\xi_N = \sqrt{N \beta (1-\beta)}$. This gives us a compact expression for the output state in Fourier space (see SM)
\begin{equation}
\label{eq:twoPhotAsymp}
|\textrm{out}\rangle_2 \sim -\frac{A \, \Gamma}{2}\int d k \, \hat{a}^{\dagger}(k) \hat{a}^{\dagger}(-k) | 0 \rangle \frac{e^{-\xi_N^2 \Gamma_{\rm tot}^2/k^2}}{k^2}.
\end{equation}
The detuned Fourier components thus dominate when $\xi_N^2 \gg 1$. The functional form of this result determines the shape of the curves in Fig.~\ref{fig:power}(a) and its inverse Fourier Transform gives the shape of $g^{(2)}(x)$ shown in Fig.~\ref{fig:g2}(b). Importantly, it also reveals that the dynamics of two-photon transport is governed universally by $\xi_N$. This is closely related to the optical depth for a resonant drive which is $\log{\left[(1-2\beta)^{2 N} \right]} \sim 4 \xi_N^2$ when $\beta\sim 0$ or $\beta \sim 1$. We highlight this in Fig.~\ref{fig:g2}(b) which shows the correlation function $\tilde{G}^{(2)}(x)= \langle \hat{a}^\dagger(0) \hat{a}^\dagger(x) \hat{a}(x) \hat{a}(0) \rangle/P_{\rm in}^2$. This is given by
\begin{equation}
\label{eq:G2Asymp}
\tilde{G}^{(2)}(x) \sim \frac{\beta^2}{4 \pi^2 \xi_N^2} \left[ \tilde{G}(\xi_N \Gamma_{\rm tot} x) \right]^2,
\end{equation}
where $\tilde{G}(x) = \int dk \, \cos{(k x)} e^{-1/k^2}/k^2$. The correlation function $\tilde{G}^{(2)}(x)$ then has the same form for all values of $\beta$ and $N$ as long as the optical depth is large. The value $\tilde{G}^{(2)}(0) = \beta^2/4 \pi \xi_N^2$ and the width of $\tilde{G}^{(2)}(x)$ scales $\propto 1/\sqrt{N}$. The correlations arising from the complex interplay between nonlinear photon interactions and dissipation can therefore be expressed in a compact universal form with a simple scaling parameter.

We now turn to the output power. This requires us to compute the contribution due to the last term in (\ref{eq:scattering}). Here we construct $\hat{S}_{12}$ by transforming from our picture of a chiral scattering process to one that contains transmission and reflection, where the coupling to the backward mode is given by our decay rate to the loss reservoir $\Gamma_{\rm tot}(1-\beta)$. In this picture we simply compute the scattering amplitude for one photon transmitted and one reflected. This is done by adapting standard two-photon scattering matrices for a single emitter (see SM) \cite{Fan2010PRA}. We do this independently for each emitter such that there are no collective effects through the loss reservoir, which is a good approximation for randomly positioned non-subwavelength emitter separations \cite{Asenjo2017PRX}. Using this scattering matrix we obtain a state which can be compactly written as
\begin{equation}
\label{eq:out21}
|\textrm{out}\rangle_{21} =\frac A 2 \sum_{M=0}^{N-1} \int dk \hat{a}^\dagger_R (k) \hat{a}^{\dagger (M+1)}_L (k) | 0 \rangle t_{k}^{N-M-1} b_M(k),
\end{equation}
where $\hat{a}^\dagger_R (k)$ and $\hat{a}^{\dagger (M+1)}_L (k)$ create right and left going photons and the superscript $M+1$ ensures there are no collective effects. The function $b_M(k_1)$ depends on $\psi_M$ and for brevity we leave its exact form for the SM. Importantly, with the state $|\textrm{out}\rangle_{21}$ at hand we can compute the power $\langle \hat{a}^{\dagger} \hat{a} \rangle $. Using our exact expressions for $\psi_N(x_c,x)$ and $b_M(k)$ we obtain an expression for $\langle \hat{a}^{\dagger} \hat{a} \rangle $ containing integrals which we evaluate numerically. These results are shown in Fig.~\ref{fig:power}(b) for different $\beta$ and $N$. Here, uncorrelated photon transport suffers exponential decay with $N$. Interestingly, we observe that the transmitted power deviates from exponential decay and for large $N$ follows a power law. The nonlinear power transmission therefore dominates for large optical depths.

We use (\ref{eq:twoPhotAsymp}) and (\ref{eq:out21})  to compute a simple asymptotic expression for the transmitted power (see SM for details)
\begin{equation}
\label{eq:powerAsymp}
\frac{\langle \hat{a}^{\dagger} \hat{a} \rangle}{P_{\rm in}}  \sim (1-2\beta)^{2N} + \frac{P_{\rm in}}{P_{\rm sat}}\frac{\beta}{4 \sqrt{\pi} \xi_N^3}\frac{3-2\beta(1-\beta)}{1-2\beta(1-\beta)},
\end{equation}
implying a nonlinear power scaling of $1/N^{3/2}$. Figure~\ref{fig:power}(b) shows excellent agreement between the full calculation and the asymptotic scaling. 
%We note that $g^{(2)}(x)$ computed to zeroth order from (\ref{eq:twoPhotAsymp}) is only valid when the linear transmission is much larger than the nonlinear term which depends on $N$, $\beta$, and $P_{\rm in}/P_{\rm sat}$. 
Finally we note that the nonlinear  power has contributions from $\hat{S}_{22}$ and $\hat{S}_{12}$, i.e., photon pairs and single photons, the nonlinear power contribution of pairs relative to single photons is $\langle \hat{a}^\dagger \hat{a} \rangle_2/\langle \hat{a}^\dagger \hat{a} \rangle_{21} \sim  1/(2\sqrt{2}-1+4\sqrt{2}/(1-2\beta(1-\beta)))$ (see SM), which is largest for $\beta \sim 0$ and $\beta \sim 1$ giving $\sim 0.13$, and smallest for $\beta=1/2$ giving $\sim 0.08$.

The physics presented here can be observed experimentally by measuring $g^{(2)}$ and $\langle \hat{a}^\dagger \hat{a} \rangle$ of the transmitted light. The nonlinear scaling of the output power and strong photon bunching are clear signals of nonlinear dynamics. State-of-the-art experimental systems that exhibit chiral light--matter interaction include quantum dots (QD) and atoms coupled to photonic nanostructures \cite{Sollner2015NNANO, Javadi2018NNANO, Mitsch2014NCOM}. In quantum dot systems the emission can be close to unidirectional and $\beta \sim 1$ \cite{Sollner2015NNANO, Mahmoodian2017OME, Javadi2018NNANO, Price2018arXiv}, however it is difficult to tune several QDs into resonance. On the other hand, hundreds of atoms can be trapped in the evanescent field of a nanofibre and exhibit chiral light-matter interaction \cite{Mitsch2014NCOM, SayrinPRX2015, Sorensen2016PRL, Corzo2016PRL, Solano2017} albeit with $\beta \ll 1$. These have a directionality of $\sim 90 \%$ and thus couple residually to the backward propagating mode, which has not been taken into account in the analytics here.

\begin{figure}[!t]
\includegraphics[width=\columnwidth]{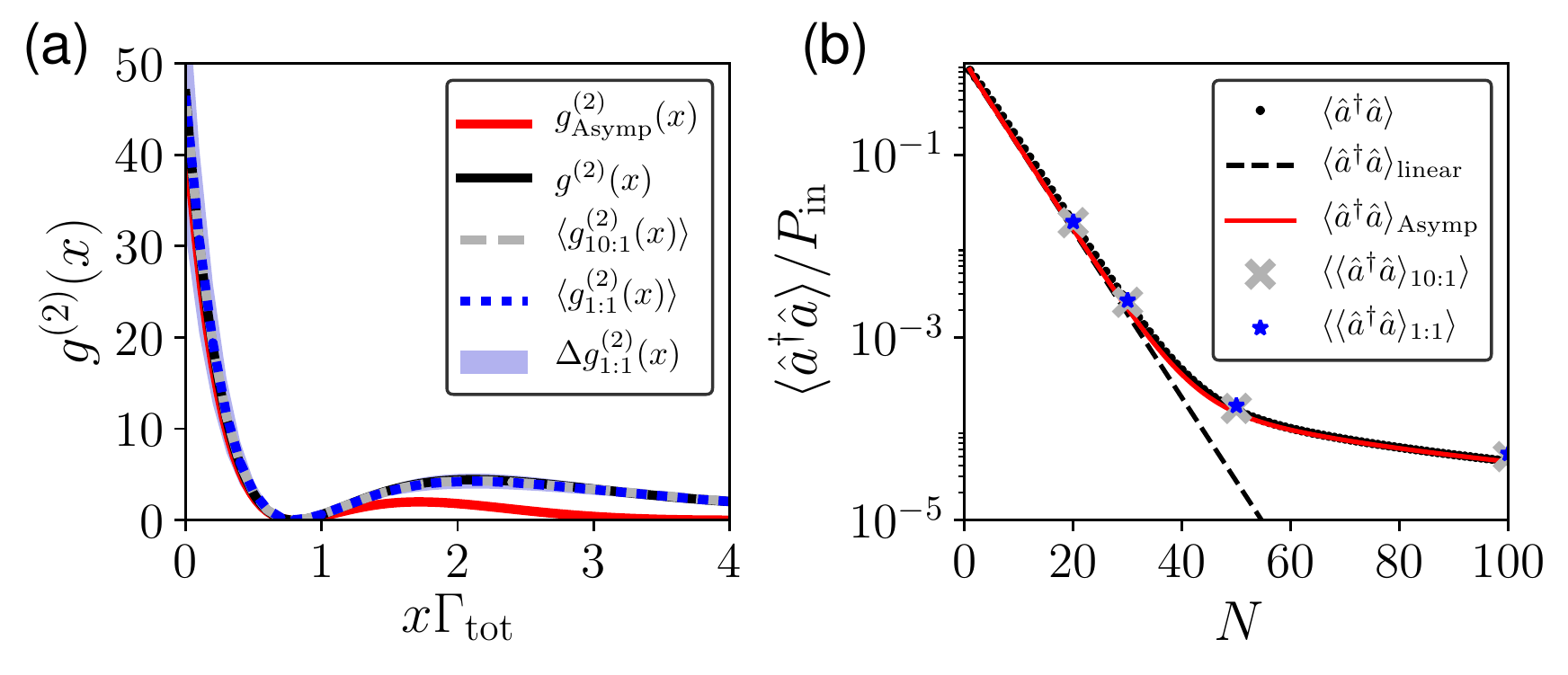}
\caption{\label{fig:bidirectional} (a) Normalized second order correlation function for $\beta=0.05$ and $N=30$. Curves show asymptotic theory $g^{(2)}_{\rm Asymp}(x)$, exact theory $g^{(2)}(x)$, and mean of numerical simulation with $\beta_L=0.005$ and $\beta_L=0.05$, $\langle g^{(2)}_{10:1}(x) \rangle$ and $\langle g^{(2)}_{1:1}(x) \rangle$ respectively. All curves, but the asymptotic theory, lie on top of another. Shading shows the standard deviation $\Delta g^{(2)}_{1:1}(x) \rangle$ for $\beta_L=\beta$. (b) Normalized output power vs emitter number for $\beta=0.05$ and $P_{\rm in}/P_{\rm sat} = 0.02$ showing the analytic theory $\langle \hat{a}^\dagger \hat{a} \rangle$, asymptotic theory $\langle \hat{a}^\dagger \hat{a} \rangle_{\rm Asymp}$, and exponential damping $\langle \hat{a}^\dagger \hat{a} \rangle_{\rm linear}$. Mean of the numerical results with $\beta_L=\beta/10$ and $\beta_L=\beta$ are $\langle \langle \hat{a}^\dagger \hat{a} \rangle_{10:1} \rangle$ and $\langle \langle \hat{a}^\dagger \hat{a} \rangle_{1:1} \rangle$ respectively.}
\end{figure}

We numerically model a system with parameters similar to a nanofiber with coupling to the forward propagating mode $\beta=0.05$ and coupling to the backward mode $\beta_L=0.005$, where the total emission rate to the waveguide is $\Gamma_{\rm tot}(\beta + \beta_L)$. We additionally model a fully bidirectional system with $\beta_L=\beta$. We use a wave function formalism where we restrict to two excitations in the system \cite{MantasCepulkovskisMastersThesis, Das2018Unpublished}, and consider $N=20,30,50,100$ emitters. The emitters are positioned randomly such that backscattering does not add up coherently. For each parameter set we consider 100 realizations. Figure \ref{fig:bidirectional}(a) shows the mean $\langle g^{(2)}(x) \rangle$ for $N=30$ when considering the ensemble. The mean agrees quantitatively with the exact unidirectional theory, while the standard deviation for $\beta_L=\beta$ shows minor discrepancies near $x \sim 0$ with standard deviation $\Delta g^{(2)}_{1:1}(0)=15$. For $\beta_L=0.005$ the standard deviation is negligible and is not shown. The asymptotic theory has a slight discrepancy because the parameters do not fall in this limit since $\xi_N^2=1.425$. Figure \ref{fig:bidirectional}(b) shows the mean output power which is also in excellent agreement with the unidirectional theory. The standard deviation of the power is insignificant on this scale and is not shown. The effects of backscattering can thus be ignored provided the number of emitters is sufficiently large and the emitters are positioned randomly.

In order to observe the physics here $P_{\rm in}$ and $N$ should be chosen such that the optical depth is sufficiently large, while the output power should be sufficiently bright to measure experimentally.  We find that for $\beta=0.05$ and $N=30$ one ideally expects a value of $g^{(2)}(0)=47$. If we consider an optical transition with $\Gamma_{\rm tot}=2\pi \times \SI{5}{\mega\hertz}$ driven with $P_{\rm in}/P_{\rm sat}=0.05$, we compute an output power of $\langle \hat{a}^\dagger \hat{a} \rangle = \SI{105}{\kilo\hertz}$ with the linear part of the power being 1.2 times larger than the nonlinear part. We also compute a coincidence rate of $\SI{1.7}{\kilo\hertz}$, where we define a coincidence as two photons separated by less than $3/\Gamma_{\rm tot}$. These outputs are sufficiently bright for detection by single photon detectors. Including the nonlinear power contribution to estimate the second order correlation for this input power gives  $g^{(2)}(0) \sim 25$. This result can be rescaled for other $\beta$ and $N$ using (\ref{eq:G2Asymp}) and (\ref{eq:powerAsymp}).

In conclusion we have analyzed the dynamics of photon--photon interactions mediated by an optically deep ensemble of emitters coupled to a waveguide. The system exhibits rich out-of-equilibrium physics due to a combination of highly nonlinear driven systems and dissipation. The emitter-induced photon--photon correlation reveals itself through the formation of bunched states of light and a universal power-law scaling of the transmission for large optical depths. As a consequence, for a sufficiently large optical depth the transmission becomes completely dominated by correlated photons. Remarkably, the formation of the strongly correlated photon states happens even for emitters weakly coupled to a waveguide and can thus be directly observed, e.g., with atoms near optical nanofibers. The present results thus open up a new avenue for studying such phenomena.

S.~M. and K.~H. acknowledge support through DFG (GRK 1991). A.~S. and S.~D. acknowledge financial support from the European Union Seventh Framework Programme through the ERC Grant QIOS (Grant No. 306576) and the Danish council for independent research (Natural Sciences). P.~L. gratefully acknowledges financial support from the European Research Council (ERC Advanced Grant ``SCALE''). P.~L. and A.~S. gratefully acknowledge financial support and the Danish National Research Foundation (Center of Excellence ``Hy-Q'').

\bibliography{bigBib}

%merlin.mbs apsrev4-1.bst 2010-07-25 4.21a (PWD, AO, DPC) hacked
%Control: key (0)
%Control: author (8) initials jnrlst
%Control: editor formatted (1) identically to author
%Control: production of article title (-1) disabled
%Control: page (0) single
%Control: year (1) truncated
%Control: production of eprint (0) enabled
\begin{thebibliography}{67}%
\makeatletter
\providecommand \@ifxundefined [1]{%
 \@ifx{#1\undefined}
}%
\providecommand \@ifnum [1]{%
 \ifnum #1\expandafter \@firstoftwo
 \else \expandafter \@secondoftwo
 \fi
}%
\providecommand \@ifx [1]{%
 \ifx #1\expandafter \@firstoftwo
 \else \expandafter \@secondoftwo
 \fi
}%
\providecommand \natexlab [1]{#1}%
\providecommand \enquote  [1]{``#1''}%
\providecommand \bibnamefont  [1]{#1}%
\providecommand \bibfnamefont [1]{#1}%
\providecommand \citenamefont [1]{#1}%
\providecommand \href@noop [0]{\@secondoftwo}%
\providecommand \href [0]{\begingroup \@sanitize@url \@href}%
\providecommand \@href[1]{\@@startlink{#1}\@@href}%
\providecommand \@@href[1]{\endgroup#1\@@endlink}%
\providecommand \@sanitize@url [0]{\catcode `\\12\catcode `\$12\catcode
  `\&12\catcode `\#12\catcode `\^12\catcode `\_12\catcode `\%12\relax}%
\providecommand \@@startlink[1]{}%
\providecommand \@@endlink[0]{}%
\providecommand \url  [0]{\begingroup\@sanitize@url \@url }%
\providecommand \@url [1]{\endgroup\@href {#1}{\urlprefix }}%
\providecommand \urlprefix  [0]{URL }%
\providecommand \Eprint [0]{\href }%
\providecommand \doibase [0]{http://dx.doi.org/}%
\providecommand \selectlanguage [0]{\@gobble}%
\providecommand \bibinfo  [0]{\@secondoftwo}%
\providecommand \bibfield  [0]{\@secondoftwo}%
\providecommand \translation [1]{[#1]}%
\providecommand \BibitemOpen [0]{}%
\providecommand \bibitemStop [0]{}%
\providecommand \bibitemNoStop [0]{.\EOS\space}%
\providecommand \EOS [0]{\spacefactor3000\relax}%
\providecommand \BibitemShut  [1]{\csname bibitem#1\endcsname}%
\let\auto@bib@innerbib\@empty
%</preamble>
\bibitem [{\citenamefont {Polkovnikov}\ \emph {et~al.}(2011)\citenamefont
  {Polkovnikov}, \citenamefont {Sengupta}, \citenamefont {Silva},\ and\
  \citenamefont {Vengalattore}}]{Polkovnikov2011RMP}%
  \BibitemOpen
  \bibfield  {author} {\bibinfo {author} {\bibfnamefont {A.}~\bibnamefont
  {Polkovnikov}}, \bibinfo {author} {\bibfnamefont {K.}~\bibnamefont
  {Sengupta}}, \bibinfo {author} {\bibfnamefont {A.}~\bibnamefont {Silva}}, \
  and\ \bibinfo {author} {\bibfnamefont {M.}~\bibnamefont {Vengalattore}},\
  }\href {\doibase 10.1103/RevModPhys.83.863} {\bibfield  {journal} {\bibinfo
  {journal} {Rev. Mod. Phys.}\ }\textbf {\bibinfo {volume} {83}},\ \bibinfo
  {pages} {863} (\bibinfo {year} {2011})}\BibitemShut {NoStop}%
\bibitem [{\citenamefont {Eisert}\ \emph {et~al.}(2015)\citenamefont {Eisert},
  \citenamefont {Friesdorf},\ and\ \citenamefont {Gogolin}}]{Eisert2015NPHYS}%
  \BibitemOpen
  \bibfield  {author} {\bibinfo {author} {\bibfnamefont {J.}~\bibnamefont
  {Eisert}}, \bibinfo {author} {\bibfnamefont {M.}~\bibnamefont {Friesdorf}}, \
  and\ \bibinfo {author} {\bibfnamefont {C.}~\bibnamefont {Gogolin}},\ }\href
  {\doibase 10.1038/nphys3215} {\bibfield  {journal} {\bibinfo  {journal} {Nat.
  Phys.}\ }\textbf {\bibinfo {volume} {11}},\ \bibinfo {pages} {124} (\bibinfo
  {year} {2015})}\BibitemShut {NoStop}%
\bibitem [{\citenamefont {Cirac}\ and\ \citenamefont
  {Zoller}(2012)}]{Cirac2012NPHYS}%
  \BibitemOpen
  \bibfield  {author} {\bibinfo {author} {\bibfnamefont {J.~I.}\ \bibnamefont
  {Cirac}}\ and\ \bibinfo {author} {\bibfnamefont {P.}~\bibnamefont {Zoller}},\
  }\href {\doibase 10.1038/nphys2275} {\bibfield  {journal} {\bibinfo
  {journal} {Nat. Phys.}\ }\textbf {\bibinfo {volume} {8}},\ \bibinfo {pages}
  {264} (\bibinfo {year} {2012})}\BibitemShut {NoStop}%
\bibitem [{\citenamefont {Datta}(1997)}]{Datta1997BOOK}%
  \BibitemOpen
  \bibfield  {author} {\bibinfo {author} {\bibfnamefont {S.}~\bibnamefont
  {Datta}},\ }\href@noop {} {\emph {\bibinfo {title} {Electronic transport in
  mesoscopic systems}}}\ (\bibinfo  {publisher} {Cambridge university press},\
  \bibinfo {year} {1997})\BibitemShut {NoStop}%
\bibitem [{\citenamefont {Das~Sarma}\ \emph {et~al.}(2011)\citenamefont
  {Das~Sarma}, \citenamefont {Adam}, \citenamefont {Hwang},\ and\ \citenamefont
  {Rossi}}]{DasSarma2011RMP}%
  \BibitemOpen
  \bibfield  {author} {\bibinfo {author} {\bibfnamefont {S.}~\bibnamefont
  {Das~Sarma}}, \bibinfo {author} {\bibfnamefont {S.}~\bibnamefont {Adam}},
  \bibinfo {author} {\bibfnamefont {E.~H.}\ \bibnamefont {Hwang}}, \ and\
  \bibinfo {author} {\bibfnamefont {E.}~\bibnamefont {Rossi}},\ }\href
  {\doibase 10.1103/RevModPhys.83.407} {\bibfield  {journal} {\bibinfo
  {journal} {Rev. Mod. Phys.}\ }\textbf {\bibinfo {volume} {83}},\ \bibinfo
  {pages} {407} (\bibinfo {year} {2011})}\BibitemShut {NoStop}%
\bibitem [{\citenamefont {Laird}\ \emph {et~al.}(2015)\citenamefont {Laird},
  \citenamefont {Kuemmeth}, \citenamefont {Steele}, \citenamefont
  {Grove-Rasmussen}, \citenamefont {Nyg\aa{}rd}, \citenamefont {Flensberg},\
  and\ \citenamefont {Kouwenhoven}}]{Laird2015RMP}%
  \BibitemOpen
  \bibfield  {author} {\bibinfo {author} {\bibfnamefont {E.~A.}\ \bibnamefont
  {Laird}}, \bibinfo {author} {\bibfnamefont {F.}~\bibnamefont {Kuemmeth}},
  \bibinfo {author} {\bibfnamefont {G.~A.}\ \bibnamefont {Steele}}, \bibinfo
  {author} {\bibfnamefont {K.}~\bibnamefont {Grove-Rasmussen}}, \bibinfo
  {author} {\bibfnamefont {J.}~\bibnamefont {Nyg\aa{}rd}}, \bibinfo {author}
  {\bibfnamefont {K.}~\bibnamefont {Flensberg}}, \ and\ \bibinfo {author}
  {\bibfnamefont {L.~P.}\ \bibnamefont {Kouwenhoven}},\ }\href {\doibase
  10.1103/RevModPhys.87.703} {\bibfield  {journal} {\bibinfo  {journal} {Rev.
  Mod. Phys.}\ }\textbf {\bibinfo {volume} {87}},\ \bibinfo {pages} {703}
  (\bibinfo {year} {2015})}\BibitemShut {NoStop}%
\bibitem [{\citenamefont {Chang}\ \emph {et~al.}(2007)\citenamefont {Chang},
  \citenamefont {S{\o}rensen}, \citenamefont {Demler},\ and\ \citenamefont
  {Lukin}}]{Chang2007NPHYS}%
  \BibitemOpen
  \bibfield  {author} {\bibinfo {author} {\bibfnamefont {D.~E.}\ \bibnamefont
  {Chang}}, \bibinfo {author} {\bibfnamefont {A.~S.}\ \bibnamefont
  {S{\o}rensen}}, \bibinfo {author} {\bibfnamefont {E.~A.}\ \bibnamefont
  {Demler}}, \ and\ \bibinfo {author} {\bibfnamefont {M.~D.}\ \bibnamefont
  {Lukin}},\ }\href {\doibase 10.1038/nphys708} {\bibfield  {journal} {\bibinfo
   {journal} {Nat. Phys.}\ }\textbf {\bibinfo {volume} {3}},\ \bibinfo {pages}
  {807} (\bibinfo {year} {2007})}\BibitemShut {NoStop}%
\bibitem [{\citenamefont {Carusotto}\ and\ \citenamefont
  {Ciuti}(2013)}]{Carusotto2013RMP}%
  \BibitemOpen
  \bibfield  {author} {\bibinfo {author} {\bibfnamefont {I.}~\bibnamefont
  {Carusotto}}\ and\ \bibinfo {author} {\bibfnamefont {C.}~\bibnamefont
  {Ciuti}},\ }\href {\doibase 10.1103/RevModPhys.85.299} {\bibfield  {journal}
  {\bibinfo  {journal} {Rev. Mod. Phys.}\ }\textbf {\bibinfo {volume} {85}},\
  \bibinfo {pages} {299} (\bibinfo {year} {2013})}\BibitemShut {NoStop}%
\bibitem [{\citenamefont {Chang}\ \emph {et~al.}(2014)\citenamefont {Chang},
  \citenamefont {Vuleti{\'c}},\ and\ \citenamefont {Lukin}}]{Chang2014NPHOT}%
  \BibitemOpen
  \bibfield  {author} {\bibinfo {author} {\bibfnamefont {D.~E.}\ \bibnamefont
  {Chang}}, \bibinfo {author} {\bibfnamefont {V.}~\bibnamefont {Vuleti{\'c}}},
  \ and\ \bibinfo {author} {\bibfnamefont {M.~D.}\ \bibnamefont {Lukin}},\
  }\href {\doibase 10.1038/nphoton.2014.192} {\bibfield  {journal} {\bibinfo
  {journal} {Nat. Photonics}\ }\textbf {\bibinfo {volume} {8}},\ \bibinfo
  {pages} {685} (\bibinfo {year} {2014})}\BibitemShut {NoStop}%
\bibitem [{\citenamefont {Dudin}\ and\ \citenamefont
  {Kuzmich}(2012)}]{Dudin2012Science}%
  \BibitemOpen
  \bibfield  {author} {\bibinfo {author} {\bibfnamefont {Y.~O.}\ \bibnamefont
  {Dudin}}\ and\ \bibinfo {author} {\bibfnamefont {A.}~\bibnamefont
  {Kuzmich}},\ }\href {\doibase 10.1126/science.1217901} {\bibfield  {journal}
  {\bibinfo  {journal} {Science}\ }\textbf {\bibinfo {volume} {336}},\ \bibinfo
  {pages} {887} (\bibinfo {year} {2012})}\BibitemShut {NoStop}%
\bibitem [{\citenamefont {Parigi}\ \emph {et~al.}(2012)\citenamefont {Parigi},
  \citenamefont {Bimbard}, \citenamefont {Stanojevic}, \citenamefont
  {Hilliard}, \citenamefont {Nogrette}, \citenamefont {Tualle-Brouri},
  \citenamefont {Ourjoumtsev},\ and\ \citenamefont {Grangier}}]{Parigi2012PRL}%
  \BibitemOpen
  \bibfield  {author} {\bibinfo {author} {\bibfnamefont {V.}~\bibnamefont
  {Parigi}}, \bibinfo {author} {\bibfnamefont {E.}~\bibnamefont {Bimbard}},
  \bibinfo {author} {\bibfnamefont {J.}~\bibnamefont {Stanojevic}}, \bibinfo
  {author} {\bibfnamefont {A.~J.}\ \bibnamefont {Hilliard}}, \bibinfo {author}
  {\bibfnamefont {F.}~\bibnamefont {Nogrette}}, \bibinfo {author}
  {\bibfnamefont {R.}~\bibnamefont {Tualle-Brouri}}, \bibinfo {author}
  {\bibfnamefont {A.}~\bibnamefont {Ourjoumtsev}}, \ and\ \bibinfo {author}
  {\bibfnamefont {P.}~\bibnamefont {Grangier}},\ }\href {\doibase
  10.1103/PhysRevLett.109.233602} {\bibfield  {journal} {\bibinfo  {journal}
  {Phys. Rev. Lett.}\ }\textbf {\bibinfo {volume} {109}},\ \bibinfo {pages}
  {233602} (\bibinfo {year} {2012})}\BibitemShut {NoStop}%
\bibitem [{\citenamefont {Peyronel}\ \emph {et~al.}(2012)\citenamefont
  {Peyronel}, \citenamefont {Firstenberg}, \citenamefont {Liang}, \citenamefont
  {Hofferberth}, \citenamefont {Gorshkov}, \citenamefont {Pohl}, \citenamefont
  {Lukin},\ and\ \citenamefont {Vuleti{\'c}}}]{Peyronel2012Nature}%
  \BibitemOpen
  \bibfield  {author} {\bibinfo {author} {\bibfnamefont {T.}~\bibnamefont
  {Peyronel}}, \bibinfo {author} {\bibfnamefont {O.}~\bibnamefont
  {Firstenberg}}, \bibinfo {author} {\bibfnamefont {Q.-Y.}\ \bibnamefont
  {Liang}}, \bibinfo {author} {\bibfnamefont {S.}~\bibnamefont {Hofferberth}},
  \bibinfo {author} {\bibfnamefont {A.~V.}\ \bibnamefont {Gorshkov}}, \bibinfo
  {author} {\bibfnamefont {T.}~\bibnamefont {Pohl}}, \bibinfo {author}
  {\bibfnamefont {M.~D.}\ \bibnamefont {Lukin}}, \ and\ \bibinfo {author}
  {\bibfnamefont {V.}~\bibnamefont {Vuleti{\'c}}},\ }\href {\doibase
  10.1038/nature11361} {\bibfield  {journal} {\bibinfo  {journal} {Nature}\
  }\textbf {\bibinfo {volume} {488}},\ \bibinfo {pages} {57} (\bibinfo {year}
  {2012})}\BibitemShut {NoStop}%
\bibitem [{\citenamefont {Maxwell}\ \emph {et~al.}(2013)\citenamefont
  {Maxwell}, \citenamefont {Szwer}, \citenamefont {Paredes-Barato},
  \citenamefont {Busche}, \citenamefont {Pritchard}, \citenamefont {Gauguet},
  \citenamefont {Weatherill}, \citenamefont {Jones},\ and\ \citenamefont
  {Adams}}]{Maxwell2013PRL}%
  \BibitemOpen
  \bibfield  {author} {\bibinfo {author} {\bibfnamefont {D.}~\bibnamefont
  {Maxwell}}, \bibinfo {author} {\bibfnamefont {D.~J.}\ \bibnamefont {Szwer}},
  \bibinfo {author} {\bibfnamefont {D.}~\bibnamefont {Paredes-Barato}},
  \bibinfo {author} {\bibfnamefont {H.}~\bibnamefont {Busche}}, \bibinfo
  {author} {\bibfnamefont {J.~D.}\ \bibnamefont {Pritchard}}, \bibinfo {author}
  {\bibfnamefont {A.}~\bibnamefont {Gauguet}}, \bibinfo {author} {\bibfnamefont
  {K.~J.}\ \bibnamefont {Weatherill}}, \bibinfo {author} {\bibfnamefont
  {M.~P.~A.}\ \bibnamefont {Jones}}, \ and\ \bibinfo {author} {\bibfnamefont
  {C.~S.}\ \bibnamefont {Adams}},\ }\href {\doibase
  10.1103/PhysRevLett.110.103001} {\bibfield  {journal} {\bibinfo  {journal}
  {Phys. Rev. Lett.}\ }\textbf {\bibinfo {volume} {110}},\ \bibinfo {pages}
  {103001} (\bibinfo {year} {2013})}\BibitemShut {NoStop}%
\bibitem [{\citenamefont {Baur}\ \emph {et~al.}(2014)\citenamefont {Baur},
  \citenamefont {Tiarks}, \citenamefont {Rempe},\ and\ \citenamefont
  {D\"urr}}]{Baur2014PRL}%
  \BibitemOpen
  \bibfield  {author} {\bibinfo {author} {\bibfnamefont {S.}~\bibnamefont
  {Baur}}, \bibinfo {author} {\bibfnamefont {D.}~\bibnamefont {Tiarks}},
  \bibinfo {author} {\bibfnamefont {G.}~\bibnamefont {Rempe}}, \ and\ \bibinfo
  {author} {\bibfnamefont {S.}~\bibnamefont {D\"urr}},\ }\href {\doibase
  10.1103/PhysRevLett.112.073901} {\bibfield  {journal} {\bibinfo  {journal}
  {Phys. Rev. Lett.}\ }\textbf {\bibinfo {volume} {112}},\ \bibinfo {pages}
  {073901} (\bibinfo {year} {2014})}\BibitemShut {NoStop}%
\bibitem [{\citenamefont {Thompson}\ \emph {et~al.}(2017)\citenamefont
  {Thompson}, \citenamefont {Nicholson}, \citenamefont {Liang}, \citenamefont
  {Cantu}, \citenamefont {Venkatramani}, \citenamefont {Choi}, \citenamefont
  {Fedorov}, \citenamefont {Viscor}, \citenamefont {Pohl}, \citenamefont
  {Lukin},\ and\ \citenamefont {Vuleti\'{c}}}]{Thompson2017Nature}%
  \BibitemOpen
  \bibfield  {author} {\bibinfo {author} {\bibfnamefont {J.~D.}\ \bibnamefont
  {Thompson}}, \bibinfo {author} {\bibfnamefont {T.~L.}\ \bibnamefont
  {Nicholson}}, \bibinfo {author} {\bibfnamefont {Q.-Y.}\ \bibnamefont
  {Liang}}, \bibinfo {author} {\bibfnamefont {S.~H.}\ \bibnamefont {Cantu}},
  \bibinfo {author} {\bibfnamefont {A.~V.}\ \bibnamefont {Venkatramani}},
  \bibinfo {author} {\bibfnamefont {S.}~\bibnamefont {Choi}}, \bibinfo {author}
  {\bibfnamefont {I.~A.}\ \bibnamefont {Fedorov}}, \bibinfo {author}
  {\bibfnamefont {D.}~\bibnamefont {Viscor}}, \bibinfo {author} {\bibfnamefont
  {T.}~\bibnamefont {Pohl}}, \bibinfo {author} {\bibfnamefont {M.~D.}\
  \bibnamefont {Lukin}}, \ and\ \bibinfo {author} {\bibfnamefont
  {V.}~\bibnamefont {Vuleti\'{c}}},\ }\href {\doibase 10.1038/nature20823}
  {\bibfield  {journal} {\bibinfo  {journal} {Nature}\ }\textbf {\bibinfo
  {volume} {542}},\ \bibinfo {pages} {206} (\bibinfo {year}
  {2017})}\BibitemShut {NoStop}%
\bibitem [{\citenamefont {Maghrebi}\ \emph {et~al.}(2015)\citenamefont
  {Maghrebi}, \citenamefont {Gullans}, \citenamefont {Bienias}, \citenamefont
  {Choi}, \citenamefont {Martin}, \citenamefont {Firstenberg}, \citenamefont
  {Lukin}, \citenamefont {B\"uchler},\ and\ \citenamefont
  {Gorshkov}}]{Maghrebi2015PRL}%
  \BibitemOpen
  \bibfield  {author} {\bibinfo {author} {\bibfnamefont {M.~F.}\ \bibnamefont
  {Maghrebi}}, \bibinfo {author} {\bibfnamefont {M.~J.}\ \bibnamefont
  {Gullans}}, \bibinfo {author} {\bibfnamefont {P.}~\bibnamefont {Bienias}},
  \bibinfo {author} {\bibfnamefont {S.}~\bibnamefont {Choi}}, \bibinfo {author}
  {\bibfnamefont {I.}~\bibnamefont {Martin}}, \bibinfo {author} {\bibfnamefont
  {O.}~\bibnamefont {Firstenberg}}, \bibinfo {author} {\bibfnamefont {M.~D.}\
  \bibnamefont {Lukin}}, \bibinfo {author} {\bibfnamefont {H.~P.}\ \bibnamefont
  {B\"uchler}}, \ and\ \bibinfo {author} {\bibfnamefont {A.~V.}\ \bibnamefont
  {Gorshkov}},\ }\href {\doibase 10.1103/PhysRevLett.115.123601} {\bibfield
  {journal} {\bibinfo  {journal} {Phys. Rev. Lett.}\ }\textbf {\bibinfo
  {volume} {115}},\ \bibinfo {pages} {123601} (\bibinfo {year}
  {2015})}\BibitemShut {NoStop}%
\bibitem [{\citenamefont {Firstenberg}\ \emph {et~al.}(2013)\citenamefont
  {Firstenberg}, \citenamefont {Peyronel}, \citenamefont {Liang}, \citenamefont
  {Gorshkov}, \citenamefont {Lukin},\ and\ \citenamefont
  {Vuleti\'{c}}}]{Firstenberg2013Nature}%
  \BibitemOpen
  \bibfield  {author} {\bibinfo {author} {\bibfnamefont {O.}~\bibnamefont
  {Firstenberg}}, \bibinfo {author} {\bibfnamefont {T.}~\bibnamefont
  {Peyronel}}, \bibinfo {author} {\bibfnamefont {Q.-Y.}\ \bibnamefont {Liang}},
  \bibinfo {author} {\bibfnamefont {A.~V.}\ \bibnamefont {Gorshkov}}, \bibinfo
  {author} {\bibfnamefont {M.~D.}\ \bibnamefont {Lukin}}, \ and\ \bibinfo
  {author} {\bibfnamefont {V.}~\bibnamefont {Vuleti\'{c}}},\ }\href {\doibase
  doi:10.1038/nature12512} {\bibfield  {journal} {\bibinfo  {journal} {Nature}\
  }\textbf {\bibinfo {volume} {502}},\ \bibinfo {pages} {71} (\bibinfo {year}
  {2013})}\BibitemShut {NoStop}%
\bibitem [{\citenamefont {Gullans}\ \emph {et~al.}(2016)\citenamefont
  {Gullans}, \citenamefont {Thompson}, \citenamefont {Wang}, \citenamefont
  {Liang}, \citenamefont {Vuleti\ifmmode~\acute{c}\else \'{c}\fi{}},
  \citenamefont {Lukin},\ and\ \citenamefont {Gorshkov}}]{Gullans2016PRL}%
  \BibitemOpen
  \bibfield  {author} {\bibinfo {author} {\bibfnamefont {M.~J.}\ \bibnamefont
  {Gullans}}, \bibinfo {author} {\bibfnamefont {J.~D.}\ \bibnamefont
  {Thompson}}, \bibinfo {author} {\bibfnamefont {Y.}~\bibnamefont {Wang}},
  \bibinfo {author} {\bibfnamefont {Q.-Y.}\ \bibnamefont {Liang}}, \bibinfo
  {author} {\bibfnamefont {V.}~\bibnamefont {Vuleti\ifmmode~\acute{c}\else
  \'{c}\fi{}}}, \bibinfo {author} {\bibfnamefont {M.~D.}\ \bibnamefont
  {Lukin}}, \ and\ \bibinfo {author} {\bibfnamefont {A.~V.}\ \bibnamefont
  {Gorshkov}},\ }\href {\doibase 10.1103/PhysRevLett.117.113601} {\bibfield
  {journal} {\bibinfo  {journal} {Phys. Rev. Lett.}\ }\textbf {\bibinfo
  {volume} {117}},\ \bibinfo {pages} {113601} (\bibinfo {year}
  {2016})}\BibitemShut {NoStop}%
\bibitem [{\citenamefont {Jachymski}\ \emph {et~al.}(2016)\citenamefont
  {Jachymski}, \citenamefont {Bienias},\ and\ \citenamefont
  {B\"uchler}}]{Jachymski2016PRL}%
  \BibitemOpen
  \bibfield  {author} {\bibinfo {author} {\bibfnamefont {K.}~\bibnamefont
  {Jachymski}}, \bibinfo {author} {\bibfnamefont {P.}~\bibnamefont {Bienias}},
  \ and\ \bibinfo {author} {\bibfnamefont {H.~P.}\ \bibnamefont {B\"uchler}},\
  }\href {\doibase 10.1103/PhysRevLett.117.053601} {\bibfield  {journal}
  {\bibinfo  {journal} {Phys. Rev. Lett.}\ }\textbf {\bibinfo {volume} {117}},\
  \bibinfo {pages} {053601} (\bibinfo {year} {2016})}\BibitemShut {NoStop}%
\bibitem [{\citenamefont {Liang}\ \emph {et~al.}(2018)\citenamefont {Liang},
  \citenamefont {Venkatramani}, \citenamefont {Cantu}, \citenamefont
  {Nicholson}, \citenamefont {Gullans}, \citenamefont {Gorshkov}, \citenamefont
  {Thompson}, \citenamefont {Chin}, \citenamefont {Lukin},\ and\ \citenamefont
  {Vuleti{\'c}}}]{Liang2017Science}%
  \BibitemOpen
  \bibfield  {author} {\bibinfo {author} {\bibfnamefont {Q.-Y.}\ \bibnamefont
  {Liang}}, \bibinfo {author} {\bibfnamefont {A.~V.}\ \bibnamefont
  {Venkatramani}}, \bibinfo {author} {\bibfnamefont {S.~H.}\ \bibnamefont
  {Cantu}}, \bibinfo {author} {\bibfnamefont {T.~L.}\ \bibnamefont
  {Nicholson}}, \bibinfo {author} {\bibfnamefont {M.~J.}\ \bibnamefont
  {Gullans}}, \bibinfo {author} {\bibfnamefont {A.~V.}\ \bibnamefont
  {Gorshkov}}, \bibinfo {author} {\bibfnamefont {J.~D.}\ \bibnamefont
  {Thompson}}, \bibinfo {author} {\bibfnamefont {C.}~\bibnamefont {Chin}},
  \bibinfo {author} {\bibfnamefont {M.~D.}\ \bibnamefont {Lukin}}, \ and\
  \bibinfo {author} {\bibfnamefont {V.}~\bibnamefont {Vuleti{\'c}}},\ }\href
  {\doibase 10.1126/science.aao7293} {\bibfield  {journal} {\bibinfo  {journal}
  {Science}\ }\textbf {\bibinfo {volume} {359}},\ \bibinfo {pages} {783}
  (\bibinfo {year} {2018})}\BibitemShut {NoStop}%
\bibitem [{\citenamefont {Goban}\ \emph {et~al.}(2014)\citenamefont {Goban},
  \citenamefont {Hung}, \citenamefont {Yu}, \citenamefont {Hood}, \citenamefont
  {Muniz}, \citenamefont {Lee}, \citenamefont {Martin}, \citenamefont
  {McClung}, \citenamefont {Choi}, \citenamefont {Chang}, \citenamefont
  {Painter},\ and\ \citenamefont {Kimble}}]{Goban2014NCOM}%
  \BibitemOpen
  \bibfield  {author} {\bibinfo {author} {\bibfnamefont {A.}~\bibnamefont
  {Goban}}, \bibinfo {author} {\bibfnamefont {C.-L.}\ \bibnamefont {Hung}},
  \bibinfo {author} {\bibfnamefont {S.-P.}\ \bibnamefont {Yu}}, \bibinfo
  {author} {\bibfnamefont {J.~D.}\ \bibnamefont {Hood}}, \bibinfo {author}
  {\bibfnamefont {J.~A.}\ \bibnamefont {Muniz}}, \bibinfo {author}
  {\bibfnamefont {J.~H.}\ \bibnamefont {Lee}}, \bibinfo {author} {\bibfnamefont
  {M.~J.}\ \bibnamefont {Martin}}, \bibinfo {author} {\bibfnamefont {A.~C.}\
  \bibnamefont {McClung}}, \bibinfo {author} {\bibfnamefont {K.~S.}\
  \bibnamefont {Choi}}, \bibinfo {author} {\bibfnamefont {D.~E.}\ \bibnamefont
  {Chang}}, \bibinfo {author} {\bibfnamefont {O.}~\bibnamefont {Painter}}, \
  and\ \bibinfo {author} {\bibfnamefont {H.~J.}\ \bibnamefont {Kimble}},\
  }\href {\doibase doi:10.1038/ncomms4808} {\bibfield  {journal} {\bibinfo
  {journal} {Nat. Commun.}\ }\textbf {\bibinfo {volume} {5}},\ \bibinfo {pages}
  {3808} (\bibinfo {year} {2014})}\BibitemShut {NoStop}%
\bibitem [{\citenamefont {Tiecke}\ \emph {et~al.}(2014)\citenamefont {Tiecke},
  \citenamefont {Thompson}, \citenamefont {de~Leon}, \citenamefont {Liu},
  \citenamefont {Vuleti\'{c}},\ and\ \citenamefont {Lukin}}]{Tiecke2014Nature}%
  \BibitemOpen
  \bibfield  {author} {\bibinfo {author} {\bibfnamefont {T.~G.}\ \bibnamefont
  {Tiecke}}, \bibinfo {author} {\bibfnamefont {J.~D.}\ \bibnamefont
  {Thompson}}, \bibinfo {author} {\bibfnamefont {N.~P.}\ \bibnamefont
  {de~Leon}}, \bibinfo {author} {\bibfnamefont {L.~R.}\ \bibnamefont {Liu}},
  \bibinfo {author} {\bibfnamefont {V.}~\bibnamefont {Vuleti\'{c}}}, \ and\
  \bibinfo {author} {\bibfnamefont {M.~D.}\ \bibnamefont {Lukin}},\ }\href
  {\doibase 10.1038/nature13188} {\bibfield  {journal} {\bibinfo  {journal}
  {Nature}\ }\textbf {\bibinfo {volume} {508}},\ \bibinfo {pages} {241}
  (\bibinfo {year} {2014})}\BibitemShut {NoStop}%
\bibitem [{\citenamefont {Hacker}\ \emph {et~al.}(2016)\citenamefont {Hacker},
  \citenamefont {Welte}, \citenamefont {Rempe},\ and\ \citenamefont
  {Ritter}}]{Hacker2016Nature}%
  \BibitemOpen
  \bibfield  {author} {\bibinfo {author} {\bibfnamefont {B.}~\bibnamefont
  {Hacker}}, \bibinfo {author} {\bibfnamefont {S.}~\bibnamefont {Welte}},
  \bibinfo {author} {\bibfnamefont {G.}~\bibnamefont {Rempe}}, \ and\ \bibinfo
  {author} {\bibfnamefont {S.}~\bibnamefont {Ritter}},\ }\href {\doibase
  doi:10.1038/nature18592} {\bibfield  {journal} {\bibinfo  {journal} {Nature}\
  }\textbf {\bibinfo {volume} {536}},\ \bibinfo {pages} {193} (\bibinfo {year}
  {2016})}\BibitemShut {NoStop}%
\bibitem [{\citenamefont {Arcari}\ \emph {et~al.}(2014)\citenamefont {Arcari},
  \citenamefont {S\"ollner}, \citenamefont {Javadi}, \citenamefont {Hansen},
  \citenamefont {Mahmoodian}, \citenamefont {Liu}, \citenamefont {Thyrrestrup},
  \citenamefont {Lee}, \citenamefont {Song}, \citenamefont {Stobbe},\ and\
  \citenamefont {Lodahl}}]{Arcari2014PRL}%
  \BibitemOpen
  \bibfield  {author} {\bibinfo {author} {\bibfnamefont {M.}~\bibnamefont
  {Arcari}}, \bibinfo {author} {\bibfnamefont {I.}~\bibnamefont {S\"ollner}},
  \bibinfo {author} {\bibfnamefont {A.}~\bibnamefont {Javadi}}, \bibinfo
  {author} {\bibfnamefont {S.~L.}\ \bibnamefont {Hansen}}, \bibinfo {author}
  {\bibfnamefont {S.}~\bibnamefont {Mahmoodian}}, \bibinfo {author}
  {\bibfnamefont {J.}~\bibnamefont {Liu}}, \bibinfo {author} {\bibfnamefont
  {H.}~\bibnamefont {Thyrrestrup}}, \bibinfo {author} {\bibfnamefont {E.~H.}\
  \bibnamefont {Lee}}, \bibinfo {author} {\bibfnamefont {J.~D.}\ \bibnamefont
  {Song}}, \bibinfo {author} {\bibfnamefont {S.}~\bibnamefont {Stobbe}}, \ and\
  \bibinfo {author} {\bibfnamefont {P.}~\bibnamefont {Lodahl}},\ }\href
  {\doibase 10.1103/PhysRevLett.113.093603} {\bibfield  {journal} {\bibinfo
  {journal} {Phys. Rev. Lett.}\ }\textbf {\bibinfo {volume} {113}},\ \bibinfo
  {pages} {093603} (\bibinfo {year} {2014})}\BibitemShut {NoStop}%
\bibitem [{\citenamefont {Lodahl}\ \emph {et~al.}(2015)\citenamefont {Lodahl},
  \citenamefont {Mahmoodian},\ and\ \citenamefont {Stobbe}}]{Lodahl2015RMP}%
  \BibitemOpen
  \bibfield  {author} {\bibinfo {author} {\bibfnamefont {P.}~\bibnamefont
  {Lodahl}}, \bibinfo {author} {\bibfnamefont {S.}~\bibnamefont {Mahmoodian}},
  \ and\ \bibinfo {author} {\bibfnamefont {S.}~\bibnamefont {Stobbe}},\ }\href
  {http://dx.doi.org/10.1103/RevModPhys.87.347} {\bibfield  {journal} {\bibinfo
   {journal} {Rev. Mod. Phys.}\ }\textbf {\bibinfo {volume} {87}},\ \bibinfo
  {pages} {347} (\bibinfo {year} {2015})}\BibitemShut {NoStop}%
\bibitem [{\citenamefont {Petersen}\ \emph {et~al.}(2014)\citenamefont
  {Petersen}, \citenamefont {Volz},\ and\ \citenamefont
  {Rauschenbeutel}}]{Petersen2014Science}%
  \BibitemOpen
  \bibfield  {author} {\bibinfo {author} {\bibfnamefont {J.}~\bibnamefont
  {Petersen}}, \bibinfo {author} {\bibfnamefont {J.}~\bibnamefont {Volz}}, \
  and\ \bibinfo {author} {\bibfnamefont {A.}~\bibnamefont {Rauschenbeutel}},\
  }\href {\doibase 10.1126/science.1257671} {\bibfield  {journal} {\bibinfo
  {journal} {Science}\ }\textbf {\bibinfo {volume} {346}},\ \bibinfo {pages}
  {67} (\bibinfo {year} {2014})}\BibitemShut {NoStop}%
\bibitem [{\citenamefont {Mitsch}\ \emph {et~al.}(2014)\citenamefont {Mitsch},
  \citenamefont {Sayrin}, \citenamefont {Albrecht}, \citenamefont
  {Schneeweiss},\ and\ \citenamefont {Rauschenbeutel}}]{Mitsch2014NCOM}%
  \BibitemOpen
  \bibfield  {author} {\bibinfo {author} {\bibfnamefont {R.}~\bibnamefont
  {Mitsch}}, \bibinfo {author} {\bibfnamefont {C.}~\bibnamefont {Sayrin}},
  \bibinfo {author} {\bibfnamefont {B.}~\bibnamefont {Albrecht}}, \bibinfo
  {author} {\bibfnamefont {P.}~\bibnamefont {Schneeweiss}}, \ and\ \bibinfo
  {author} {\bibfnamefont {A.}~\bibnamefont {Rauschenbeutel}},\ }\href
  {\doibase 10.1038/ncomms6713} {\bibfield  {journal} {\bibinfo  {journal}
  {Nat. Commun.}\ }\textbf {\bibinfo {volume} {5}},\ \bibinfo {pages} {5713}
  (\bibinfo {year} {2014})}\BibitemShut {NoStop}%
\bibitem [{\citenamefont {S\"{o}llner}\ \emph {et~al.}(2015)\citenamefont
  {S\"{o}llner}, \citenamefont {Mahmoodian}, \citenamefont {Hansen},
  \citenamefont {Midolo}, \citenamefont {Kirsanske}, \citenamefont
  {Pregnolato}, \citenamefont {El-Ella}, \citenamefont {Lee}, \citenamefont
  {Song}, \citenamefont {Stobbe},\ and\ \citenamefont
  {Lodahl}}]{Sollner2015NNANO}%
  \BibitemOpen
  \bibfield  {author} {\bibinfo {author} {\bibfnamefont {I.}~\bibnamefont
  {S\"{o}llner}}, \bibinfo {author} {\bibfnamefont {S.}~\bibnamefont
  {Mahmoodian}}, \bibinfo {author} {\bibfnamefont {S.~L.}\ \bibnamefont
  {Hansen}}, \bibinfo {author} {\bibfnamefont {L.}~\bibnamefont {Midolo}},
  \bibinfo {author} {\bibfnamefont {G.}~\bibnamefont {Kirsanske}}, \bibinfo
  {author} {\bibfnamefont {T.}~\bibnamefont {Pregnolato}}, \bibinfo {author}
  {\bibfnamefont {H.}~\bibnamefont {El-Ella}}, \bibinfo {author} {\bibfnamefont
  {E.~H.}\ \bibnamefont {Lee}}, \bibinfo {author} {\bibfnamefont {J.~D.}\
  \bibnamefont {Song}}, \bibinfo {author} {\bibfnamefont {S.}~\bibnamefont
  {Stobbe}}, \ and\ \bibinfo {author} {\bibfnamefont {P.}~\bibnamefont
  {Lodahl}},\ }\href {\doibase 10.1038/nnano.2015.159} {\bibfield  {journal}
  {\bibinfo  {journal} {Nat. Nanotechnol.}\ }\textbf {\bibinfo {volume} {10}},\
  \bibinfo {pages} {775} (\bibinfo {year} {2015})}\BibitemShut {NoStop}%
\bibitem [{\citenamefont {Young}\ \emph {et~al.}(2015)\citenamefont {Young},
  \citenamefont {Thijssen}, \citenamefont {Beggs}, \citenamefont
  {Androvitsaneas}, \citenamefont {Kuipers}, \citenamefont {Rarity},
  \citenamefont {Hughes},\ and\ \citenamefont {Oulton}}]{Young2015PRL}%
  \BibitemOpen
  \bibfield  {author} {\bibinfo {author} {\bibfnamefont {A.~B.}\ \bibnamefont
  {Young}}, \bibinfo {author} {\bibfnamefont {A.~C.~T.}\ \bibnamefont
  {Thijssen}}, \bibinfo {author} {\bibfnamefont {D.~M.}\ \bibnamefont {Beggs}},
  \bibinfo {author} {\bibfnamefont {P.}~\bibnamefont {Androvitsaneas}},
  \bibinfo {author} {\bibfnamefont {L.}~\bibnamefont {Kuipers}}, \bibinfo
  {author} {\bibfnamefont {J.~G.}\ \bibnamefont {Rarity}}, \bibinfo {author}
  {\bibfnamefont {S.}~\bibnamefont {Hughes}}, \ and\ \bibinfo {author}
  {\bibfnamefont {R.}~\bibnamefont {Oulton}},\ }\href {\doibase
  10.1103/PhysRevLett.115.153901} {\bibfield  {journal} {\bibinfo  {journal}
  {Phys. Rev. Lett.}\ }\textbf {\bibinfo {volume} {115}},\ \bibinfo {pages}
  {153901} (\bibinfo {year} {2015})}\BibitemShut {NoStop}%
\bibitem [{\citenamefont {Coles}\ \emph {et~al.}(2016)\citenamefont {Coles},
  \citenamefont {Price}, \citenamefont {Dixon}, \citenamefont {Royall},
  \citenamefont {Clarke}, \citenamefont {Kok}, \citenamefont {Skolnick},
  \citenamefont {Fox},\ and\ \citenamefont {Makhonin}}]{Coles2016NCOM}%
  \BibitemOpen
  \bibfield  {author} {\bibinfo {author} {\bibfnamefont {R.~J.}\ \bibnamefont
  {Coles}}, \bibinfo {author} {\bibfnamefont {D.~M.}\ \bibnamefont {Price}},
  \bibinfo {author} {\bibfnamefont {J.~E.}\ \bibnamefont {Dixon}}, \bibinfo
  {author} {\bibfnamefont {B.}~\bibnamefont {Royall}}, \bibinfo {author}
  {\bibfnamefont {E.}~\bibnamefont {Clarke}}, \bibinfo {author} {\bibfnamefont
  {P.}~\bibnamefont {Kok}}, \bibinfo {author} {\bibfnamefont {M.~S.}\
  \bibnamefont {Skolnick}}, \bibinfo {author} {\bibfnamefont {A.~M.}\
  \bibnamefont {Fox}}, \ and\ \bibinfo {author} {\bibfnamefont {M.~N.}\
  \bibnamefont {Makhonin}},\ }\href {\doibase 10.1038/ncomms11183} {\bibfield
  {journal} {\bibinfo  {journal} {Nat. Commun.}\ }\textbf {\bibinfo {volume}
  {7}},\ \bibinfo {pages} {11183} (\bibinfo {year} {2016})}\BibitemShut
  {NoStop}%
\bibitem [{\citenamefont {Lodahl}\ \emph {et~al.}(2017)\citenamefont {Lodahl},
  \citenamefont {Mahmoodian}, \citenamefont {Stobbe}, \citenamefont
  {Rauschenbeutel}, \citenamefont {Schneeweiss}, \citenamefont {Volz},
  \citenamefont {Pichler},\ and\ \citenamefont {Zoller}}]{Lodahl2017Nature}%
  \BibitemOpen
  \bibfield  {author} {\bibinfo {author} {\bibfnamefont {P.}~\bibnamefont
  {Lodahl}}, \bibinfo {author} {\bibfnamefont {S.}~\bibnamefont {Mahmoodian}},
  \bibinfo {author} {\bibfnamefont {S.}~\bibnamefont {Stobbe}}, \bibinfo
  {author} {\bibfnamefont {A.}~\bibnamefont {Rauschenbeutel}}, \bibinfo
  {author} {\bibfnamefont {P.}~\bibnamefont {Schneeweiss}}, \bibinfo {author}
  {\bibfnamefont {J.}~\bibnamefont {Volz}}, \bibinfo {author} {\bibfnamefont
  {H.}~\bibnamefont {Pichler}}, \ and\ \bibinfo {author} {\bibfnamefont
  {P.}~\bibnamefont {Zoller}},\ }\href {\doibase doi:10.1038/nature21037}
  {\bibfield  {journal} {\bibinfo  {journal} {Nature}\ }\textbf {\bibinfo
  {volume} {541}},\ \bibinfo {pages} {473} (\bibinfo {year}
  {2017})}\BibitemShut {NoStop}%
\bibitem [{\citenamefont {Stannigel}\ \emph {et~al.}(2012)\citenamefont
  {Stannigel}, \citenamefont {Rabl},\ and\ \citenamefont
  {Zoller}}]{Stannigel2012NJP}%
  \BibitemOpen
  \bibfield  {author} {\bibinfo {author} {\bibfnamefont {K.}~\bibnamefont
  {Stannigel}}, \bibinfo {author} {\bibfnamefont {P.}~\bibnamefont {Rabl}}, \
  and\ \bibinfo {author} {\bibfnamefont {P.}~\bibnamefont {Zoller}},\ }\href
  {http://stacks.iop.org/1367-2630/14/i=6/a=063014} {\bibfield  {journal}
  {\bibinfo  {journal} {New J. Phys.}\ }\textbf {\bibinfo {volume} {14}},\
  \bibinfo {pages} {063014} (\bibinfo {year} {2012})}\BibitemShut {NoStop}%
\bibitem [{\citenamefont {Ramos}\ \emph {et~al.}(2014)\citenamefont {Ramos},
  \citenamefont {Pichler}, \citenamefont {Daley},\ and\ \citenamefont
  {Zoller}}]{Ramos2014quantum}%
  \BibitemOpen
  \bibfield  {author} {\bibinfo {author} {\bibfnamefont {T.}~\bibnamefont
  {Ramos}}, \bibinfo {author} {\bibfnamefont {H.}~\bibnamefont {Pichler}},
  \bibinfo {author} {\bibfnamefont {A.~J.}\ \bibnamefont {Daley}}, \ and\
  \bibinfo {author} {\bibfnamefont {P.}~\bibnamefont {Zoller}},\ }\href
  {\doibase 10.1103/PhysRevLett.113.237203} {\bibfield  {journal} {\bibinfo
  {journal} {Phys. Rev. Lett.}\ }\textbf {\bibinfo {volume} {113}},\ \bibinfo
  {pages} {237203} (\bibinfo {year} {2014})}\BibitemShut {NoStop}%
\bibitem [{\citenamefont {Pichler}\ \emph {et~al.}(2015)\citenamefont
  {Pichler}, \citenamefont {Ramos}, \citenamefont {Daley},\ and\ \citenamefont
  {Zoller}}]{Pichler2015PRA}%
  \BibitemOpen
  \bibfield  {author} {\bibinfo {author} {\bibfnamefont {H.}~\bibnamefont
  {Pichler}}, \bibinfo {author} {\bibfnamefont {T.}~\bibnamefont {Ramos}},
  \bibinfo {author} {\bibfnamefont {A.~J.}\ \bibnamefont {Daley}}, \ and\
  \bibinfo {author} {\bibfnamefont {P.}~\bibnamefont {Zoller}},\ }\href
  {\doibase 10.1103/PhysRevA.91.042116} {\bibfield  {journal} {\bibinfo
  {journal} {Phys. Rev. A}\ }\textbf {\bibinfo {volume} {91}},\ \bibinfo
  {pages} {042116} (\bibinfo {year} {2015})}\BibitemShut {NoStop}%
\bibitem [{\citenamefont {Mahmoodian}\ \emph {et~al.}(2016)\citenamefont
  {Mahmoodian}, \citenamefont {Lodahl},\ and\ \citenamefont
  {S\o{}rensen}}]{Mahmoodian2016PRL}%
  \BibitemOpen
  \bibfield  {author} {\bibinfo {author} {\bibfnamefont {S.}~\bibnamefont
  {Mahmoodian}}, \bibinfo {author} {\bibfnamefont {P.}~\bibnamefont {Lodahl}},
  \ and\ \bibinfo {author} {\bibfnamefont {A.~S.}\ \bibnamefont
  {S\o{}rensen}},\ }\href {\doibase 10.1103/PhysRevLett.117.240501} {\bibfield
  {journal} {\bibinfo  {journal} {Phys. Rev. Lett.}\ }\textbf {\bibinfo
  {volume} {117}},\ \bibinfo {pages} {240501} (\bibinfo {year}
  {2016})}\BibitemShut {NoStop}%
\bibitem [{\citenamefont {Scheucher}\ \emph {et~al.}(2016)\citenamefont
  {Scheucher}, \citenamefont {Hilico}, \citenamefont {Will}, \citenamefont
  {Volz},\ and\ \citenamefont {Rauschenbeutel}}]{Scheucher2016Science}%
  \BibitemOpen
  \bibfield  {author} {\bibinfo {author} {\bibfnamefont {M.}~\bibnamefont
  {Scheucher}}, \bibinfo {author} {\bibfnamefont {A.}~\bibnamefont {Hilico}},
  \bibinfo {author} {\bibfnamefont {E.}~\bibnamefont {Will}}, \bibinfo {author}
  {\bibfnamefont {J.}~\bibnamefont {Volz}}, \ and\ \bibinfo {author}
  {\bibfnamefont {A.}~\bibnamefont {Rauschenbeutel}},\ }\href {\doibase
  10.1126/science.aaj2118} {\bibfield  {journal} {\bibinfo  {journal}
  {Science}\ }\textbf {\bibinfo {volume} {354}},\ \bibinfo {pages} {1577}
  (\bibinfo {year} {2016})}\BibitemShut {NoStop}%
\bibitem [{\citenamefont {Zeuthen}\ \emph {et~al.}(2017)\citenamefont
  {Zeuthen}, \citenamefont {Gullans}, \citenamefont {Maghrebi},\ and\
  \citenamefont {Gorshkov}}]{Zeuthen2017PRL}%
  \BibitemOpen
  \bibfield  {author} {\bibinfo {author} {\bibfnamefont {E.}~\bibnamefont
  {Zeuthen}}, \bibinfo {author} {\bibfnamefont {M.~J.}\ \bibnamefont
  {Gullans}}, \bibinfo {author} {\bibfnamefont {M.~F.}\ \bibnamefont
  {Maghrebi}}, \ and\ \bibinfo {author} {\bibfnamefont {A.~V.}\ \bibnamefont
  {Gorshkov}},\ }\href {\doibase 10.1103/PhysRevLett.119.043602} {\bibfield
  {journal} {\bibinfo  {journal} {Phys. Rev. Lett.}\ }\textbf {\bibinfo
  {volume} {119}},\ \bibinfo {pages} {043602} (\bibinfo {year}
  {2017})}\BibitemShut {NoStop}%
\bibitem [{\citenamefont {Murray}\ \emph {et~al.}(2018)\citenamefont {Murray},
  \citenamefont {Mirgorodskiy}, \citenamefont {Tresp}, \citenamefont {Braun},
  \citenamefont {Paris-Mandoki}, \citenamefont {Gorshkov}, \citenamefont
  {Hofferberth},\ and\ \citenamefont {Pohl}}]{Murray2018PRL}%
  \BibitemOpen
  \bibfield  {author} {\bibinfo {author} {\bibfnamefont {C.~R.}\ \bibnamefont
  {Murray}}, \bibinfo {author} {\bibfnamefont {I.}~\bibnamefont
  {Mirgorodskiy}}, \bibinfo {author} {\bibfnamefont {C.}~\bibnamefont {Tresp}},
  \bibinfo {author} {\bibfnamefont {C.}~\bibnamefont {Braun}}, \bibinfo
  {author} {\bibfnamefont {A.}~\bibnamefont {Paris-Mandoki}}, \bibinfo {author}
  {\bibfnamefont {A.~V.}\ \bibnamefont {Gorshkov}}, \bibinfo {author}
  {\bibfnamefont {S.}~\bibnamefont {Hofferberth}}, \ and\ \bibinfo {author}
  {\bibfnamefont {T.}~\bibnamefont {Pohl}},\ }\href {\doibase
  10.1103/PhysRevLett.120.113601} {\bibfield  {journal} {\bibinfo  {journal}
  {Phys. Rev. Lett.}\ }\textbf {\bibinfo {volume} {120}},\ \bibinfo {pages}
  {113601} (\bibinfo {year} {2018})}\BibitemShut {NoStop}%
\bibitem [{\citenamefont {Sayrin}\ \emph {et~al.}(2015)\citenamefont {Sayrin},
  \citenamefont {Junge}, \citenamefont {Mitsch}, \citenamefont {Albrecht},
  \citenamefont {O'Shea}, \citenamefont {Schneeweiss}, \citenamefont {Volz},\
  and\ \citenamefont {Rauschenbeutel}}]{SayrinPRX2015}%
  \BibitemOpen
  \bibfield  {author} {\bibinfo {author} {\bibfnamefont {C.}~\bibnamefont
  {Sayrin}}, \bibinfo {author} {\bibfnamefont {C.}~\bibnamefont {Junge}},
  \bibinfo {author} {\bibfnamefont {R.}~\bibnamefont {Mitsch}}, \bibinfo
  {author} {\bibfnamefont {B.}~\bibnamefont {Albrecht}}, \bibinfo {author}
  {\bibfnamefont {D.}~\bibnamefont {O'Shea}}, \bibinfo {author} {\bibfnamefont
  {P.}~\bibnamefont {Schneeweiss}}, \bibinfo {author} {\bibfnamefont
  {J.}~\bibnamefont {Volz}}, \ and\ \bibinfo {author} {\bibfnamefont
  {A.}~\bibnamefont {Rauschenbeutel}},\ }\href {\doibase
  10.1103/PhysRevX.5.041036} {\bibfield  {journal} {\bibinfo  {journal} {Phys.
  Rev. X}\ }\textbf {\bibinfo {volume} {5}},\ \bibinfo {pages} {041036}
  (\bibinfo {year} {2015})}\BibitemShut {NoStop}%
\bibitem [{\citenamefont {S\o{}rensen}\ \emph {et~al.}(2016)\citenamefont
  {S\o{}rensen}, \citenamefont {B\'eguin}, \citenamefont {Kluge}, \citenamefont
  {Iakoupov}, \citenamefont {S\o{}rensen}, \citenamefont {M\"uller},
  \citenamefont {Polzik},\ and\ \citenamefont {Appel}}]{Sorensen2016PRL}%
  \BibitemOpen
  \bibfield  {author} {\bibinfo {author} {\bibfnamefont {H.~L.}\ \bibnamefont
  {S\o{}rensen}}, \bibinfo {author} {\bibfnamefont {J.-B.}\ \bibnamefont
  {B\'eguin}}, \bibinfo {author} {\bibfnamefont {K.~W.}\ \bibnamefont {Kluge}},
  \bibinfo {author} {\bibfnamefont {I.}~\bibnamefont {Iakoupov}}, \bibinfo
  {author} {\bibfnamefont {A.~S.}\ \bibnamefont {S\o{}rensen}}, \bibinfo
  {author} {\bibfnamefont {J.~H.}\ \bibnamefont {M\"uller}}, \bibinfo {author}
  {\bibfnamefont {E.~S.}\ \bibnamefont {Polzik}}, \ and\ \bibinfo {author}
  {\bibfnamefont {J.}~\bibnamefont {Appel}},\ }\href {\doibase
  10.1103/PhysRevLett.117.133604} {\bibfield  {journal} {\bibinfo  {journal}
  {Phys. Rev. Lett.}\ }\textbf {\bibinfo {volume} {117}},\ \bibinfo {pages}
  {133604} (\bibinfo {year} {2016})}\BibitemShut {NoStop}%
\bibitem [{\citenamefont {Corzo}\ \emph {et~al.}(2016)\citenamefont {Corzo},
  \citenamefont {Gouraud}, \citenamefont {Chandra}, \citenamefont {Goban},
  \citenamefont {Sheremet}, \citenamefont {Kupriyanov},\ and\ \citenamefont
  {Laurat}}]{Corzo2016PRL}%
  \BibitemOpen
  \bibfield  {author} {\bibinfo {author} {\bibfnamefont {N.~V.}\ \bibnamefont
  {Corzo}}, \bibinfo {author} {\bibfnamefont {B.}~\bibnamefont {Gouraud}},
  \bibinfo {author} {\bibfnamefont {A.}~\bibnamefont {Chandra}}, \bibinfo
  {author} {\bibfnamefont {A.}~\bibnamefont {Goban}}, \bibinfo {author}
  {\bibfnamefont {A.~S.}\ \bibnamefont {Sheremet}}, \bibinfo {author}
  {\bibfnamefont {D.~V.}\ \bibnamefont {Kupriyanov}}, \ and\ \bibinfo {author}
  {\bibfnamefont {J.}~\bibnamefont {Laurat}},\ }\href {\doibase
  10.1103/PhysRevLett.117.133603} {\bibfield  {journal} {\bibinfo  {journal}
  {Phys. Rev. Lett.}\ }\textbf {\bibinfo {volume} {117}},\ \bibinfo {pages}
  {133603} (\bibinfo {year} {2016})}\BibitemShut {NoStop}%
\bibitem [{\citenamefont {Solano}\ \emph {et~al.}(2017)\citenamefont {Solano},
  \citenamefont {Grover}, \citenamefont {Hoffman}, \citenamefont {Ravets},
  \citenamefont {Fatemi}, \citenamefont {Orozco},\ and\ \citenamefont
  {Rolston}}]{Solano2017}%
  \BibitemOpen
  \bibfield  {author} {\bibinfo {author} {\bibfnamefont {P.}~\bibnamefont
  {Solano}}, \bibinfo {author} {\bibfnamefont {J.~A.}\ \bibnamefont {Grover}},
  \bibinfo {author} {\bibfnamefont {J.~E.}\ \bibnamefont {Hoffman}}, \bibinfo
  {author} {\bibfnamefont {S.}~\bibnamefont {Ravets}}, \bibinfo {author}
  {\bibfnamefont {F.~K.}\ \bibnamefont {Fatemi}}, \bibinfo {author}
  {\bibfnamefont {L.~A.}\ \bibnamefont {Orozco}}, \ and\ \bibinfo {author}
  {\bibfnamefont {S.~L.}\ \bibnamefont {Rolston}},\ }in\ \href {\doibase
  10.1016/bs.aamop.2017.02.003} {\emph {\bibinfo {booktitle} {Advances In
  Atomic, Molecular, and Optical Physics}}},\ Vol.~\bibinfo {volume} {66}\
  (\bibinfo  {publisher} {Elsevier},\ \bibinfo {year} {2017})\ pp.\ \bibinfo
  {pages} {439--505}\BibitemShut {NoStop}%
\bibitem [{\citenamefont {Stanley}(1971)}]{StanleyBOOK}%
  \BibitemOpen
  \bibfield  {author} {\bibinfo {author} {\bibfnamefont {H.~E.}\ \bibnamefont
  {Stanley}},\ }\href@noop {} {\emph {\bibinfo {title} {Phase transitions and
  critical phenomena}}}\ (\bibinfo  {publisher} {Clarendon Press, Oxford},\
  \bibinfo {year} {1971})\BibitemShut {NoStop}%
\bibitem [{\citenamefont {Thompson}\ and\ \citenamefont
  {Stewart}(2002)}]{ThompsonBOOK}%
  \BibitemOpen
  \bibfield  {author} {\bibinfo {author} {\bibfnamefont {J.~M.~T.}\
  \bibnamefont {Thompson}}\ and\ \bibinfo {author} {\bibfnamefont {H.~B.}\
  \bibnamefont {Stewart}},\ }\href@noop {} {\emph {\bibinfo {title} {Nonlinear
  dynamics and chaos}}}\ (\bibinfo  {publisher} {John Wiley \& Sons},\ \bibinfo
  {year} {2002})\BibitemShut {NoStop}%
\bibitem [{\citenamefont {Carmichael}(1993)}]{Carmichael1993PRL}%
  \BibitemOpen
  \bibfield  {author} {\bibinfo {author} {\bibfnamefont {H.~J.}\ \bibnamefont
  {Carmichael}},\ }\href {\doibase 10.1103/PhysRevLett.70.2273} {\bibfield
  {journal} {\bibinfo  {journal} {Phys. Rev. Lett.}\ }\textbf {\bibinfo
  {volume} {70}},\ \bibinfo {pages} {2273} (\bibinfo {year}
  {1993})}\BibitemShut {NoStop}%
\bibitem [{\citenamefont {Gardiner}(1993)}]{Gardiner1993PRL}%
  \BibitemOpen
  \bibfield  {author} {\bibinfo {author} {\bibfnamefont {C.~W.}\ \bibnamefont
  {Gardiner}},\ }\href {\doibase 10.1103/PhysRevLett.70.2269} {\bibfield
  {journal} {\bibinfo  {journal} {Phys. Rev. Lett.}\ }\textbf {\bibinfo
  {volume} {70}},\ \bibinfo {pages} {2269} (\bibinfo {year}
  {1993})}\BibitemShut {NoStop}%
\bibitem [{\citenamefont {Asenjo-Garcia}\ \emph
  {et~al.}(2017{\natexlab{a}})\citenamefont {Asenjo-Garcia}, \citenamefont
  {Hood}, \citenamefont {Chang},\ and\ \citenamefont {Kimble}}]{Asenjo2017PRA}%
  \BibitemOpen
  \bibfield  {author} {\bibinfo {author} {\bibfnamefont {A.}~\bibnamefont
  {Asenjo-Garcia}}, \bibinfo {author} {\bibfnamefont {J.~D.}\ \bibnamefont
  {Hood}}, \bibinfo {author} {\bibfnamefont {D.~E.}\ \bibnamefont {Chang}}, \
  and\ \bibinfo {author} {\bibfnamefont {H.~J.}\ \bibnamefont {Kimble}},\
  }\href {\doibase 10.1103/PhysRevA.95.033818} {\bibfield  {journal} {\bibinfo
  {journal} {Phys. Rev. A}\ }\textbf {\bibinfo {volume} {95}},\ \bibinfo
  {pages} {033818} (\bibinfo {year} {2017}{\natexlab{a}})}\BibitemShut
  {NoStop}%
\bibitem [{\citenamefont {Manzoni}\ \emph {et~al.}()\citenamefont {Manzoni},
  \citenamefont {Chang},\ and\ \citenamefont {Douglas}}]{Manzoni2017NCOM}%
  \BibitemOpen
  \bibfield  {author} {\bibinfo {author} {\bibfnamefont {M.~T.}\ \bibnamefont
  {Manzoni}}, \bibinfo {author} {\bibfnamefont {D.~E.}\ \bibnamefont {Chang}},
  \ and\ \bibinfo {author} {\bibfnamefont {J.~S.}\ \bibnamefont {Douglas}},\
  }\href {\doibase 10.1038/s41467-017-01416-4} {\bibfield  {journal} {\bibinfo
  {journal} {Nat. Commun.}\ }\textbf {\bibinfo {volume} {8}},\ \bibinfo {pages}
  {1743}}\BibitemShut {NoStop}%
\bibitem [{\citenamefont {Shen}\ and\ \citenamefont {Fan}(2005)}]{Shen2005OL}%
  \BibitemOpen
  \bibfield  {author} {\bibinfo {author} {\bibfnamefont {J.~T.}\ \bibnamefont
  {Shen}}\ and\ \bibinfo {author} {\bibfnamefont {S.}~\bibnamefont {Fan}},\
  }\href {\doibase 10.1364/OL.30.002001} {\bibfield  {journal} {\bibinfo
  {journal} {Opt. Lett.}\ }\textbf {\bibinfo {volume} {30}},\ \bibinfo {pages}
  {2001} (\bibinfo {year} {2005})}\BibitemShut {NoStop}%
\bibitem [{\citenamefont {Kojima}\ \emph {et~al.}(2003)\citenamefont {Kojima},
  \citenamefont {Hofmann}, \citenamefont {Takeuchi},\ and\ \citenamefont
  {Sasaki}}]{Kojima2003PRA}%
  \BibitemOpen
  \bibfield  {author} {\bibinfo {author} {\bibfnamefont {K.}~\bibnamefont
  {Kojima}}, \bibinfo {author} {\bibfnamefont {H.~F.}\ \bibnamefont {Hofmann}},
  \bibinfo {author} {\bibfnamefont {S.}~\bibnamefont {Takeuchi}}, \ and\
  \bibinfo {author} {\bibfnamefont {K.}~\bibnamefont {Sasaki}},\ }\href
  {\doibase 10.1103/PhysRevA.68.013803} {\bibfield  {journal} {\bibinfo
  {journal} {Phys. Rev. A}\ }\textbf {\bibinfo {volume} {68}},\ \bibinfo
  {pages} {013803} (\bibinfo {year} {2003})}\BibitemShut {NoStop}%
\bibitem [{\citenamefont {Hofmann}\ \emph {et~al.}(2003)\citenamefont
  {Hofmann}, \citenamefont {Kojima}, \citenamefont {Takeuchi},\ and\
  \citenamefont {Sasaki}}]{Hofmann2003PRA}%
  \BibitemOpen
  \bibfield  {author} {\bibinfo {author} {\bibfnamefont {H.~F.}\ \bibnamefont
  {Hofmann}}, \bibinfo {author} {\bibfnamefont {K.}~\bibnamefont {Kojima}},
  \bibinfo {author} {\bibfnamefont {S.}~\bibnamefont {Takeuchi}}, \ and\
  \bibinfo {author} {\bibfnamefont {K.}~\bibnamefont {Sasaki}},\ }\href
  {\doibase 10.1103/PhysRevA.68.043813} {\bibfield  {journal} {\bibinfo
  {journal} {Phys. Rev. A}\ }\textbf {\bibinfo {volume} {68}},\ \bibinfo
  {pages} {043813} (\bibinfo {year} {2003})}\BibitemShut {NoStop}%
\bibitem [{\citenamefont {Shen}\ and\ \citenamefont {Fan}(2007)}]{Shen2007PRA}%
  \BibitemOpen
  \bibfield  {author} {\bibinfo {author} {\bibfnamefont {J.-T.}\ \bibnamefont
  {Shen}}\ and\ \bibinfo {author} {\bibfnamefont {S.}~\bibnamefont {Fan}},\
  }\href {\doibase 10.1103/PhysRevA.76.062709} {\bibfield  {journal} {\bibinfo
  {journal} {Phys. Rev. A}\ }\textbf {\bibinfo {volume} {76}},\ \bibinfo
  {pages} {062709} (\bibinfo {year} {2007})}\BibitemShut {NoStop}%
\bibitem [{\citenamefont {Zheng}\ \emph {et~al.}(2010)\citenamefont {Zheng},
  \citenamefont {Gauthier},\ and\ \citenamefont {Baranger}}]{Zheng2010PRA}%
  \BibitemOpen
  \bibfield  {author} {\bibinfo {author} {\bibfnamefont {H.}~\bibnamefont
  {Zheng}}, \bibinfo {author} {\bibfnamefont {D.~J.}\ \bibnamefont {Gauthier}},
  \ and\ \bibinfo {author} {\bibfnamefont {H.~U.}\ \bibnamefont {Baranger}},\
  }\href {\doibase 10.1103/PhysRevA.82.063816} {\bibfield  {journal} {\bibinfo
  {journal} {Phys. Rev. A}\ }\textbf {\bibinfo {volume} {82}},\ \bibinfo
  {pages} {063816} (\bibinfo {year} {2010})}\BibitemShut {NoStop}%
\bibitem [{\citenamefont {Fan}\ \emph {et~al.}(2010)\citenamefont {Fan},
  \citenamefont {\c{S}\"{u}kr\"{u} Ekin~Kocaba\c{s}},\ and\ \citenamefont
  {Shen}}]{Fan2010PRA}%
  \BibitemOpen
  \bibfield  {author} {\bibinfo {author} {\bibfnamefont {S.}~\bibnamefont
  {Fan}}, \bibinfo {author} {\bibnamefont {\c{S}\"{u}kr\"{u}
  Ekin~Kocaba\c{s}}}, \ and\ \bibinfo {author} {\bibfnamefont {J.-T.}\
  \bibnamefont {Shen}},\ }\href {\doibase 10.1103/PhysRevA.82.063821}
  {\bibfield  {journal} {\bibinfo  {journal} {Phys. Rev. A}\ }\textbf {\bibinfo
  {volume} {82}},\ \bibinfo {pages} {063821} (\bibinfo {year}
  {2010})}\BibitemShut {NoStop}%
\bibitem [{\citenamefont {Shi}\ \emph {et~al.}(2015)\citenamefont {Shi},
  \citenamefont {Chang},\ and\ \citenamefont {Cirac}}]{Shi2015PRA}%
  \BibitemOpen
  \bibfield  {author} {\bibinfo {author} {\bibfnamefont {T.}~\bibnamefont
  {Shi}}, \bibinfo {author} {\bibfnamefont {D.~E.}\ \bibnamefont {Chang}}, \
  and\ \bibinfo {author} {\bibfnamefont {J.~I.}\ \bibnamefont {Cirac}},\ }\href
  {\doibase 10.1103/PhysRevA.92.053834} {\bibfield  {journal} {\bibinfo
  {journal} {Phys. Rev. A}\ }\textbf {\bibinfo {volume} {92}},\ \bibinfo
  {pages} {053834} (\bibinfo {year} {2015})}\BibitemShut {NoStop}%
\bibitem [{\citenamefont {Rupasov}(1982)}]{Rupasov1982JETP}%
  \BibitemOpen
  \bibfield  {author} {\bibinfo {author} {\bibfnamefont {V.~I.}\ \bibnamefont
  {Rupasov}},\ }\href@noop {} {\bibfield  {journal} {\bibinfo  {journal} {JETP
  Lett.}\ }\textbf {\bibinfo {volume} {36}},\ \bibinfo {pages} {142} (\bibinfo
  {year} {1982})}\BibitemShut {NoStop}%
\bibitem [{\citenamefont {Rupasov}\ and\ \citenamefont
  {Yudson}(1984)}]{Rupasov1984JETP}%
  \BibitemOpen
  \bibfield  {author} {\bibinfo {author} {\bibfnamefont {V.~I.}\ \bibnamefont
  {Rupasov}}\ and\ \bibinfo {author} {\bibfnamefont {V.~I.}\ \bibnamefont
  {Yudson}},\ }\href@noop {} {\bibfield  {journal} {\bibinfo  {journal} {Sov.
  Phys. JETP}\ }\textbf {\bibinfo {volume} {59}},\ \bibinfo {pages} {478}
  (\bibinfo {year} {1984})}\BibitemShut {NoStop}%
\bibitem [{\citenamefont {Yudson}(1985)}]{Yudson1985JETP}%
  \BibitemOpen
  \bibfield  {author} {\bibinfo {author} {\bibfnamefont {V.~I.}\ \bibnamefont
  {Yudson}},\ }\href@noop {} {\bibfield  {journal} {\bibinfo  {journal} {Sov.
  Phys. JETP}\ }\textbf {\bibinfo {volume} {88}},\ \bibinfo {pages} {1043}
  (\bibinfo {year} {1985})}\BibitemShut {NoStop}%
\bibitem [{\citenamefont {Pletyukhov}\ and\ \citenamefont
  {Gritsev}(2012)}]{Pletyukhov2012NJP}%
  \BibitemOpen
  \bibfield  {author} {\bibinfo {author} {\bibfnamefont {M.}~\bibnamefont
  {Pletyukhov}}\ and\ \bibinfo {author} {\bibfnamefont {V.}~\bibnamefont
  {Gritsev}},\ }\href {\doibase 10.1088/1367-2630/14/9/095028} {\bibfield
  {journal} {\bibinfo  {journal} {New J. Phys.}\ }\textbf {\bibinfo {volume}
  {14}},\ \bibinfo {pages} {095028} (\bibinfo {year} {2012})}\BibitemShut
  {NoStop}%
\bibitem [{\citenamefont {Ringel}\ \emph {et~al.}(2014)\citenamefont {Ringel},
  \citenamefont {Pletyukhov},\ and\ \citenamefont {Gritsev}}]{Ringel2014NJP}%
  \BibitemOpen
  \bibfield  {author} {\bibinfo {author} {\bibfnamefont {M.}~\bibnamefont
  {Ringel}}, \bibinfo {author} {\bibfnamefont {M.}~\bibnamefont {Pletyukhov}},
  \ and\ \bibinfo {author} {\bibfnamefont {V.}~\bibnamefont {Gritsev}},\ }\href
  {\doibase 10.1088/1367-2630/16/11/113030} {\bibfield  {journal} {\bibinfo
  {journal} {New J. Phys.}\ }\textbf {\bibinfo {volume} {16}},\ \bibinfo
  {pages} {113030} (\bibinfo {year} {2014})}\BibitemShut {NoStop}%
\bibitem [{\citenamefont {Chang}\ \emph {et~al.}(2012)\citenamefont {Chang},
  \citenamefont {Jiang}, \citenamefont {Gorshkov},\ and\ \citenamefont
  {Kimble}}]{Chang2012NJP}%
  \BibitemOpen
  \bibfield  {author} {\bibinfo {author} {\bibfnamefont {D.~E.}\ \bibnamefont
  {Chang}}, \bibinfo {author} {\bibfnamefont {L.}~\bibnamefont {Jiang}},
  \bibinfo {author} {\bibfnamefont {A.~V.}\ \bibnamefont {Gorshkov}}, \ and\
  \bibinfo {author} {\bibfnamefont {H.~J.}\ \bibnamefont {Kimble}},\ }\href
  {\doibase 10.1088/1367-2630/14/6/063003} {\bibfield  {journal} {\bibinfo
  {journal} {New J. Phys.}\ }\textbf {\bibinfo {volume} {14}},\ \bibinfo
  {pages} {063003} (\bibinfo {year} {2012})}\BibitemShut {NoStop}%
\bibitem [{\citenamefont {Asenjo-Garcia}\ \emph
  {et~al.}(2017{\natexlab{b}})\citenamefont {Asenjo-Garcia}, \citenamefont
  {Moreno-Cardoner}, \citenamefont {Albrecht}, \citenamefont {Kimble},\ and\
  \citenamefont {Chang}}]{Asenjo2017PRX}%
  \BibitemOpen
  \bibfield  {author} {\bibinfo {author} {\bibfnamefont {A.}~\bibnamefont
  {Asenjo-Garcia}}, \bibinfo {author} {\bibfnamefont {M.}~\bibnamefont
  {Moreno-Cardoner}}, \bibinfo {author} {\bibfnamefont {A.}~\bibnamefont
  {Albrecht}}, \bibinfo {author} {\bibfnamefont {H.~J.}\ \bibnamefont
  {Kimble}}, \ and\ \bibinfo {author} {\bibfnamefont {D.~E.}\ \bibnamefont
  {Chang}},\ }\href {\doibase 10.1103/PhysRevX.7.031024} {\bibfield  {journal}
  {\bibinfo  {journal} {Phys. Rev. X}\ }\textbf {\bibinfo {volume} {7}},\
  \bibinfo {pages} {031024} (\bibinfo {year} {2017}{\natexlab{b}})}\BibitemShut
  {NoStop}%
\bibitem [{\citenamefont {Javadi}\ \emph {et~al.}(2018)\citenamefont {Javadi},
  \citenamefont {Ding}, \citenamefont {Appel}, \citenamefont {Mahmoodian},
  \citenamefont {L\"{o}bl}, \citenamefont {S\"{o}llner}, \citenamefont
  {Schott}, \citenamefont {Papon}, \citenamefont {Pregnolato}, \citenamefont
  {Stobbe}, \citenamefont {Midolo}, \citenamefont {Schr\"{o}der}, \citenamefont
  {Wieck}, \citenamefont {Ludwig}, \citenamefont {Warburton},\ and\
  \citenamefont {Lodahl}}]{Javadi2018NNANO}%
  \BibitemOpen
  \bibfield  {author} {\bibinfo {author} {\bibfnamefont {A.}~\bibnamefont
  {Javadi}}, \bibinfo {author} {\bibfnamefont {D.}~\bibnamefont {Ding}},
  \bibinfo {author} {\bibfnamefont {M.~H.}\ \bibnamefont {Appel}}, \bibinfo
  {author} {\bibfnamefont {S.}~\bibnamefont {Mahmoodian}}, \bibinfo {author}
  {\bibfnamefont {M.~C.}\ \bibnamefont {L\"{o}bl}}, \bibinfo {author}
  {\bibfnamefont {I.}~\bibnamefont {S\"{o}llner}}, \bibinfo {author}
  {\bibfnamefont {R.}~\bibnamefont {Schott}}, \bibinfo {author} {\bibfnamefont
  {C.}~\bibnamefont {Papon}}, \bibinfo {author} {\bibfnamefont
  {T.}~\bibnamefont {Pregnolato}}, \bibinfo {author} {\bibfnamefont
  {S.}~\bibnamefont {Stobbe}}, \bibinfo {author} {\bibfnamefont
  {L.}~\bibnamefont {Midolo}}, \bibinfo {author} {\bibfnamefont
  {T.}~\bibnamefont {Schr\"{o}der}}, \bibinfo {author} {\bibfnamefont {A.~D.}\
  \bibnamefont {Wieck}}, \bibinfo {author} {\bibfnamefont {A.}~\bibnamefont
  {Ludwig}}, \bibinfo {author} {\bibfnamefont {R.~J.}\ \bibnamefont
  {Warburton}}, \ and\ \bibinfo {author} {\bibfnamefont {P.}~\bibnamefont
  {Lodahl}},\ }\href {\doibase 10.1038/s41565-018-0091-5} {\bibfield  {journal}
  {\bibinfo  {journal} {Nat. Nanotechnol.}\ }\textbf {\bibinfo {volume} {13}},\
  \bibinfo {pages} {398} (\bibinfo {year} {2018})}\BibitemShut {NoStop}%
\bibitem [{\citenamefont {Mahmoodian}\ \emph {et~al.}(2017)\citenamefont
  {Mahmoodian}, \citenamefont {Prindal-Nielsen}, \citenamefont {S\"{o}llner},
  \citenamefont {Stobbe},\ and\ \citenamefont {Lodahl}}]{Mahmoodian2017OME}%
  \BibitemOpen
  \bibfield  {author} {\bibinfo {author} {\bibfnamefont {S.}~\bibnamefont
  {Mahmoodian}}, \bibinfo {author} {\bibfnamefont {K.}~\bibnamefont
  {Prindal-Nielsen}}, \bibinfo {author} {\bibfnamefont {I.}~\bibnamefont
  {S\"{o}llner}}, \bibinfo {author} {\bibfnamefont {S.}~\bibnamefont {Stobbe}},
  \ and\ \bibinfo {author} {\bibfnamefont {P.}~\bibnamefont {Lodahl}},\ }\href
  {\doibase 10.1364/OME.7.000043} {\bibfield  {journal} {\bibinfo  {journal}
  {Opt. Mater. Express}\ }\textbf {\bibinfo {volume} {7}},\ \bibinfo {pages}
  {43} (\bibinfo {year} {2017})}\BibitemShut {NoStop}%
\bibitem [{\citenamefont {Price}\ \emph {et~al.}(2018)\citenamefont {Price},
  \citenamefont {Hurst}, \citenamefont {Bentham}, \citenamefont {Makhonin},
  \citenamefont {Royall}, \citenamefont {Clarke}, \citenamefont {Kok},
  \citenamefont {Wilson}, \citenamefont {Skolnick},\ and\ \citenamefont
  {Fox}}]{Price2018arXiv}%
  \BibitemOpen
  \bibfield  {author} {\bibinfo {author} {\bibfnamefont {D.~M.}\ \bibnamefont
  {Price}}, \bibinfo {author} {\bibfnamefont {D.~L.}\ \bibnamefont {Hurst}},
  \bibinfo {author} {\bibfnamefont {C.}~\bibnamefont {Bentham}}, \bibinfo
  {author} {\bibfnamefont {M.~N.}\ \bibnamefont {Makhonin}}, \bibinfo {author}
  {\bibfnamefont {B.}~\bibnamefont {Royall}}, \bibinfo {author} {\bibfnamefont
  {E.}~\bibnamefont {Clarke}}, \bibinfo {author} {\bibfnamefont
  {P.}~\bibnamefont {Kok}}, \bibinfo {author} {\bibfnamefont {L.}~\bibnamefont
  {Wilson}}, \bibinfo {author} {\bibfnamefont {M.}~\bibnamefont {Skolnick}}, \
  and\ \bibinfo {author} {\bibfnamefont {A.}~\bibnamefont {Fox}},\ }\href@noop
  {} {\bibfield  {journal} {\bibinfo  {journal} {arXiv:1801.09958}\ } (\bibinfo
  {year} {2018})}\BibitemShut {NoStop}%
\bibitem [{\citenamefont
  {\v{C}epulkovskis}(2017)}]{MantasCepulkovskisMastersThesis}%
  \BibitemOpen
  \bibfield  {author} {\bibinfo {author} {\bibfnamefont {M.}~\bibnamefont
  {\v{C}epulkovskis}},\ }\emph {\bibinfo {title} {Nonlinear photon interactions
  in waveguides}},\ \href@noop {} {Master's thesis},\ \bibinfo  {school} {Niels
  Bohr Institute, University of Copenhagen} (\bibinfo {year}
  {2017})\BibitemShut {NoStop}%
\bibitem [{\citenamefont {Das}(2018)}]{Das2018Unpublished}%
  \BibitemOpen
  \bibfield  {author} {\bibinfo {author} {\bibfnamefont {S.}~\bibnamefont
  {Das}},\ }\href@noop {} {\bibfield  {journal} {\bibinfo  {journal} {In
  preparation}\ } (\bibinfo {year} {2018})}\BibitemShut {NoStop}%
\end{thebibliography}%

%\appendix
\onecolumngrid
\newpage

%%%%%%%%%% Merge with supplemental materials %%%%%%%%%%
\pagebreak
\widetext
\begin{center}
\textbf{\large Supplementary Material: Chiral waveguide QED: Strongly correlated photon transport with weakly coupled emitters}
\end{center}
%%%%%%%%%% Merge with supplemental materials %%%%%%%%%%
%%%%%%%%%% Prefix a "S" to all equations, figures, sections, tables and reset the counter %%%%%%%%%%
\setcounter{equation}{0}
\setcounter{figure}{0}
\setcounter{table}{0}
\setcounter{page}{1}
\makeatletter
\renewcommand{\theequation}{S\arabic{equation}}
\renewcommand{\thefigure}{S\arabic{figure}}
\renewcommand{\thesection}{S\arabic{section}}
%\renewcommand{\bibnumfmt}[1]{[S#1]}
%\renewcommand{\citenumfont}[1]{S#1}
%%%%%%%%%% Prefix a "S" to all equations, figures, tables and reset the counter %%%%%%%%%%

\section{Computing $\hat{S}_{22}$}

In this section we introduce the scattering matrix $\hat{S}_{22}$ and compute the output state $|\textrm{out} \rangle_{2}$. We start from the input state
\begin{equation}
\begin{split}
|\alpha_{\rm in} \rangle &= e^{-\frac{|\alpha|^2}{2}}\left[1 + \alpha \hat{a}^\dagger_{k_0} + \frac{\alpha^2}{2} \hat{a}^\dagger_{k_0} \hat{a}^\dagger_{k_0}  + \ldots \right] | 0 \rangle \\
&= e^{-\frac{|\alpha|^2}{2}}\left[1 + \sqrt{\frac{2\pi}{L}}\alpha \hat{a}^\dagger (k_0) + \frac{2\pi}{L} \frac{\alpha^2}{2} \hat{a}^\dagger(k_0) \hat{a}^\dagger(k_0) + \ldots \right] | 0 \rangle, \\
\end{split}
\end{equation}
where $L$ is a quantization length, and in the second line we have switched from single-mode operators to operators $\hat{a}^\dagger_{k_0} \rightarrow \sqrt{\frac{2\pi}{L}} \hat{a}^\dagger(k_0)$ suitable for taking the continuum limit $L \rightarrow \infty$. The two-photon scattering matrix in Ref.~\cite{Shen2007PRA} can be easily generalized to $N$ chirally coupled emitters giving
\begin{equation}
\begin{split}
\hat{S}_{22}^N = \frac 1 2 \int \, dE d\Delta t_{\frac E 2 + \Delta}^N t_{\frac E 2 - \Delta}^N |W_{E, \Delta} \rangle \langle W_{E,\Delta} | + \int dE \, \tilde{t}_{E}^N | B_E \rangle \langle B_E |,
\end{split}
\end{equation}
where all integrals range from $-\infty$ to $\infty$, and the two-photon scattering eigenstates in a position-space representation are
\begin{equation}
\begin{split}
|W_{E,\Delta} \rangle &= \frac{1}{\sqrt{2}}\int dx_1 dx_2  \hat{a}^\dagger({x_1}) \hat{a}^\dagger ({x_2})| 0 \rangle \, W_{E,\Delta} (x_c,x)\\
|B_{E} \rangle &= \frac{1}{\sqrt{2}}\int dx_1 dx_2  \hat{a}^\dagger({x_1}) \hat{a}^\dagger ({x_2})| 0 \rangle \, B_{E} (x_c,x).
\end{split}
\end{equation}
Here,
\begin{equation}
\begin{split}
W_{E,\Delta} (x_c,x) &= \frac{1}{\sqrt{4 \Delta^2 + \Gamma^2}}\frac{\sqrt{2}}{2 \pi}e^{i E x_c}\left[2 \Delta \cos{(\Delta x)} - \Gamma \operatorname{sgn}{(x)}\sin{(\Delta x)} \right],\\
B_{E} (x_c,x) &= \sqrt{\frac{\Gamma}{4\pi}}\,e^{i E x_c}\,e^{-\frac \Gamma 2 |x|},\\
\end{split}
\end{equation}
and
\begin{equation}
\begin{split}
t_k &= \frac{k + i \,\Gamma (1-2\beta)/(2\beta)}{k + i \Gamma/(2\beta)}\\
\tilde{t}_E &= \frac{E + i \Gamma (1-3\beta)/\beta}{E + i \Gamma(1+\beta)/\beta},\\
\end{split}
\end{equation}
and $x_c=\frac{x_1+x_2}{2}$, $x=x_1-x_2$, $E=k+p$ is a two-photon detuning, and $\Delta=\frac{k-p}{2}$ is a difference in photon energies, where $k$ and $p$ are photon detunings. The eigenstates $|W_{E,\Delta} \rangle$ and $|B_{E} \rangle$ form an orthonormal basis for the two-photon subspace \cite{Shen2007PRA}. Projecting the input state on the two-photon scattering matrix $|\textrm{out}\rangle_2 \equiv \hat{S}_{22}^N | \alpha_{\rm in}\rangle $ gives
\begin{equation}
\begin{split}
\label{eq:out2X}
|\textrm{out}\rangle_2 &= \frac A 2 \int dx_1 dx_2 \hat{a}^\dagger(x_1) \hat{a}^\dagger(x_2) | 0 \rangle \psi_N(x_c,x),\\
\psi_N(x_c,x)&= 2 \, \tilde{t}_{2k_0}^N e^{2 i k_0 x_c}e^{- \frac{\Gamma}{2} |x|} - \frac{\Gamma}{\pi} e^{2 i k_0 x_c} \int d \Delta \frac{t_{k_0+\Delta}^N t_{k_0-\Delta}^N}{\Gamma^2 + 4 \Delta^2}\left[ 2 \cos{(\Delta x) -\frac \Gamma \Delta} \operatorname{sgn}{(x)} \sin{(\Delta x)} \right].
\end{split}
\end{equation}
These integrals can be computed analytically. For details see the section below. Combining the terms together we obtain
\begin{equation}
\begin{split}
\label{eq:psiDiffer}
&\psi_N(x_c,x) = e^{2i k_0 x_c} \left\{ t_{k_0}^{2N} -  \frac{i\Gamma/2}{(N-1)!} \frac{d^{N-1}}{dz^{N-1}} \left[ \frac{t_{k_0 + z}^N (z-k_0 - i\Gamma (1-2\beta)/2\beta)^N e^{i z |x|}}{z^2 + \Gamma^2/4}\right]_{z=k_0 + \frac{i \Gamma}{2 \beta}} \right.\\
& + \frac{i\Gamma/2}{(N-1)!} \frac{d^{N-1}}{dz^{N-1}} \left[ \frac{t_{k_0 - z}^N (z+k_0 +i\Gamma (1-2\beta)/2\beta)^N e^{-i z |x|}}{z^2 + \Gamma^2/4}\right]_{z=-k_0 - \frac{i \Gamma}{2 \beta}}\\
&+ \frac{\Gamma^2/4}{(N-1)!} \frac{d^{N-1}}{dz^{N-1}} \left[ \frac{t_{k_0 + z}^N (z-k_0 - i\Gamma (1-2\beta)/2\beta)^N e^{i z |x|}}{z(z^2 + \Gamma^2/4)}\right]_{z=k_0 + \frac{i \Gamma}{2 \beta}}\\
&+ \left. \frac{\Gamma^2/4}{(N-1)!} \frac{d^{N-1}}{dz^{N-1}} \left[ \frac{t_{k_0 - z}^N (z+k_0 +i\Gamma (1-2\beta)/2\beta)^N e^{-i z |x|}}{z(z^2 + \Gamma^2/4)}\right]_{z=-k_0 - \frac{i \Gamma}{2 \beta}}\right\},
\end{split}
\end{equation}
which expresses the solution in terms of an $(N-1)$th order differential operator. These differential operators can be evaluated in terms of generalized functions (see section below). Using the results leading up to (\ref{eq:psiExpression}) we express the two-photon wavefunction as
\begin{equation}
\label{eq:psiFull}
\psi_N(x_c,x) = e^{2i k_0 x_c}\left\lbrace t_{k_0}^{2N} - \frac{1}{(N-1)!}\sum_{n=0}^{N-1} \binom{N-1}{n} F_{k_0}(N,n) \chi_{k_0,N-1-n}(x) \right\rbrace,
\end{equation}
where $F_{k_0}(N,n)$ is given by (\ref{eq:3F2}) and $\chi_{k_0,n}(x)$ by (\ref{eq:chiExpression}). This is defined in the main text as $\psi_N(x_c,x) = e^{2i k_0 x_c}\left[ t_{k_0}^{2N} - \phi_N(x) \right] $.

It is also useful to express the two-photon output state in $k$-space. We simply rewrite
(\ref{eq:out2X}) in terms of creation operators $\hat{a}^\dagger(k_1)$ and $\hat{a}^\dagger(k_2)$
\begin{equation}
\begin{split}
\label{eq:out2K}
|\textrm{out}\rangle_2 &= \frac A 2 \int dk_1 dk_2 \hat{a}^\dagger(k_1) \hat{a}^\dagger(k_2) | 0 \rangle \psi_N(E_k, \Delta_k),\\
\end{split}
\end{equation}
where $E_k = k_1 + k_2$ and $\Delta_k = (k_1 - k_2)/2$ and
\begin{equation}
\psi_N(E_k, \Delta_k) = \delta(E_k - 2k_0) \left\lbrace 2\pi \delta(\Delta_k)t_{k_0}^{2 N} - \frac{1}{(N-1)!}\sum_{n=0}^{N-1}\binom{N-1}{n} F_{k_0}(N,n) \chi_{N-1-n,k_0}(\Delta_k) \right \rbrace.
\end{equation}
We have
\begin{equation}
\chi_{n,k_0}(\Delta_k) = n! \, \Gamma \left\lbrace \frac{ \left[-\Delta_k  - \gamma \right]^{-n-1}}{\Delta_k(\Delta_k + \frac{i \Gamma}{2})}  + \frac{ \left[\Delta_k  - \gamma \right]^{-n-1}}{\Delta_k(\Delta_k - \frac{i \Gamma}{2})}   - \frac{2}{\Delta_k^2 +\frac{\Gamma^2}{4}} \left(\frac{-1}{a_0}\right)^{n+1}\right\rbrace,
\end{equation}
$\gamma = k_0 + i\Gamma/2\beta$, and $a_0=k_0 + i\Gamma(1-\beta)/2\beta$. We define in a compact form $\psi_N(E_k,\Delta_k) = \delta(E_k - 2k_0)\left[2\pi \delta(\Delta_k)t_{2k_0}^{2N} - \phi_N(\Delta_k) \right]$.

\section{Constructing and computing $\hat{S}_{12}$}

\begin{figure}[!t]
\includegraphics[width=0.8\columnwidth]{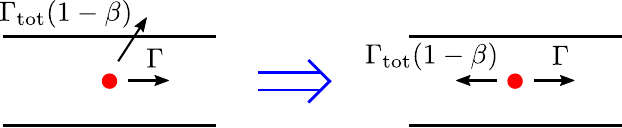}
\caption{\label{fig:S21} For a single emitter we can map a unidirectional system with losses to a bidirectional system. We can use this to compute the multi-emitter scattering matrix $\hat{S}_{12}$ by ignoring collective effects.}
\end{figure}

In this section we compute the term due to one photon being scattered out of the waveguide and the other being transmitted. Each of the $N$ emitters contributes to this term and we therefore have to sum over all the emitters. We start by applying the scattering matrix in Eq.~(\ref{eq:scattering}) on the input state and obtain the $k$-space result for the two-photon output state after scattering off $M$ emitters,
\begin{equation}
\label{eq:S21start}
2\pi \frac A 2 \sum_{M=0}^{N-1}\hat{S}_{11}^{N-M-1} \hat{S}_{12} \hat{S}_{22}^M \hat{a}^\dagger (k_0) \hat{a}^\dagger (k_0) | 0 \rangle = \frac A 2  \sum_{M=0}^{N-1}\hat{S}_{11}^{N-M-1} \hat{S}_{12} \int dk_1 dk_2 \, \psi_M(E_k,\Delta_k) \, \hat{a}^\dagger (k_1) \hat{a}^\dagger (k_2) | 0 \rangle.
\end{equation}
The task at hand is thus to apply $\hat{S}_{12}$ to each term in the sum.

We construct $\hat{S}_{12}$ by mapping a chiral interaction with losses to a bidirectional waveguide where the emission rate to the backward propagating mode is the same as the loss rate in the chiral system as illustrated in Fig.~\ref{fig:S21}. From this we can construct $\hat{S}_{12}$ from the scattering matrix for one photon transmitted and one reflected by adapting the process outlined in Ref.~\cite{Fan2010PRA}. This scattering matrix for an arbitrary two-photon wavefunction $f(k_1,k_2)$ is
\begin{equation}
\begin{split}
\hat{S}_{12} \int dk_1 dk_2 \hat{a}^\dagger(k_1) \hat{a}^\dagger(k_2) | 0 \rangle f(k_1,k_2) = \int dk_1 dk_2 \hat{a}_R^\dagger(k_1) \hat{a}_L^\dagger(-k_2) | 0 \rangle \left[ 2 \bar{r}_{-k_2}t_{k_1} f(k_1,-k_2) \vphantom{\frac{i\beta\sqrt{\Gamma}}{\pi}}\right.\\
\left.+ \frac{i\beta\sqrt{\Gamma(1-\beta)}}{\pi}\bar{s}_{k_1} \bar{s}_{-k_2}  \int dp_1 dp_2 (\bar{s}_{p_1} + \bar{s}_{p_2}) \delta(k_1-k_2-p_1-p_2) f(p_1,p_2)\right],
\end{split}
\end{equation}
where $\hat{a}_R^\dagger$ creates a forward-propagating photon and $\hat{a}_L^\dagger$ creates a backward propagating photon, and backwards photons have negative wavevectors. Here, $\bar{r}_k = -2i\sqrt{\beta(1-\beta)}/(1-2 i k/\Gamma_{\rm tot})$ and $\bar{s}_k = \sqrt{\Gamma_{\rm tot}}/(k + i \Gamma_{\rm tot}/2)$, where we have used the overbar as these definitions differ from those typically used in the literature. Applying $\hat{S}_{12}$ and $S_{11}$ in (\ref{eq:S21start}) gives
\begin{equation}
\begin{split}
\label{eq:out21BIG}
|\textrm{out}\rangle_{21} = \frac A 2  \sum_{M=0}^{N-1} \int dk_1 dk_2 \, t_{k_1}^{N-M-1} \, \hat{a}_R^\dagger (k_1) \hat{a}_L^{\dagger (M+1)} (-k_2) | 0 \rangle \left[ 2 \, \bar{r}_{-k_2} t_{k_1} \, \psi_M\left(2\Delta_k,\frac{E_k}{2} \right) + \frac{i\beta\sqrt{\Gamma (1-\beta)}}{\pi}\right.\\
\left. \vphantom{\frac{\beta\sqrt{\Gamma}}{\pi}}\times \bar{s}_{k_1} \bar{s}_{-k_2} \int d E_p d\Delta_p \, \left(\bar{s}_{\frac{E_p}{2}+\Delta_p} + \bar{s}_{\frac{E_p}{2}-\Delta_p} \right) \, \delta(2\Delta_k - E_p) \, \psi_M(E_p,\Delta_p)\right],
\end{split}
\end{equation}
where the superscript $(M+1)$ on $\hat{a}^{\dagger}$ ensures each emitter is coupled to a separate reservoir such that the loss reservoir does not mediate collective effects. Considering a resonant drive $k_0 \rightarrow 0$ and using the expression for $\psi_N(E_k,\Delta_k)$ in (\ref{eq:out2K}) after some manipulation we obtain
\begin{equation}
\begin{split}
|\textrm{out}\rangle_{21} &= \frac A 2  \sum_{M=0}^{N-1} \int dk \, t_{k}^{N-M-1} \, \hat{a}_R^\dagger (k) \hat{a}_L^{\dagger (M+1)} (-k) | 0 \rangle \left\lbrace 4 \pi \, \bar{r}_{-k} t_{k} t_0^{2M} \delta(k) - 2 \bar{r}_{-k}t_{k} \phi_M(k) \vphantom{\frac{\sqrt{\Gamma}}{\pi}} \right.\\
&\left. + \frac{i\beta\sqrt{\Gamma (1-\beta)}}{\pi} \bar{s}_{k} \bar{s}_{-k} \left[ 4 \pi \bar{s}_0 t_0^{2M} - \Phi_M \right] \right\rbrace\\
&\equiv \frac A 2  \sum_{M=0}^{N-1} \int dk \, t_{k}^{N-M-1} \, \hat{a}_R^\dagger (k) \hat{a}_L^{\dagger (M+1)} (-k) | 0 \rangle b_M(k),
\end{split}
\end{equation}
which gives (\ref{eq:out21}) in the main text. Here, we also define 
\begin{equation}
\label{eq:c_M}
c_M(k) = \left\lbrace - 2 \bar{r}_{-k}t_{k} \phi_M(k) + \frac{i\beta\sqrt{\Gamma (1-\beta)}}{\pi} \bar{s}_{k} \bar{s}_{-k} \left[ 4 \pi \bar{s}_0 t_0^{2M} - \Phi_M \right] \right\rbrace,
\end{equation}
which is the correlated part of $|\textrm{out}\rangle_{21}$. The term $\Phi_M = \int dk \, (\bar{s}_k + \bar{s}_{-k}) \phi_M(k)$ can be computed analytically (see section below).

\section{Output power}

In this section we compute the power output after $N$ emitters for a resonant drive. The terms $| \textrm{out}\rangle_1$, $| \textrm{out}\rangle_2$, and $| \textrm{out}\rangle_{21}$ all contribute to this, and we write $\langle \hat{a}^\dagger \hat{a} \rangle = \langle \hat{a}^\dagger \hat{a} \rangle_1 + \langle \hat{a}^\dagger \hat{a} \rangle_2 + \langle \hat{a}^\dagger \hat{a} \rangle_{21}$. We note that the system is at steady state so we evaluate the power at $x=0$ without loss of generality.  Computing $\langle \hat{a}^\dagger \hat{a} \rangle_1$ is straightforward and gives $\langle \hat{a}^\dagger \hat{a} \rangle_1 = \frac{|\alpha|^2 e^{-|\alpha|^2}}{L}t_{0}^{2N} \sim t_0^{2N}(|\alpha|^2/L - |\alpha|^4 L/L^2)$, where $t_0=1-2\beta$. The term $|\alpha|^4 L/L^2$ is non-physical as it scales with $L$ (after taking into account that the input  power is $P_{\rm in}= |\alpha|^2/L$). We below show that it cancels with similar unphysical terms from the two-photon contribution.

Moving to $\langle \hat{a}^\dagger \hat{a} \rangle_2$, we compute $\hat{a}(x=0) |\textrm{out} \rangle_2$ and use this to find
\begin{equation}
\langle \hat{a}^\dagger \hat{a} \rangle_2 = \frac{e^{-|\alpha|^2}|\alpha|^4}{L^2} \left[ |t_0|^{4N} L + \int dx |\phi_N (x)|^2 - 2 \, t_0^{2N} \int dx \phi_N(x) \right].
\end{equation}
Again, the first term here is non-physical as it depends on $L$. This term cancels with the similar non-physical term from $\langle \hat{a}^\dagger \hat{a} \rangle_{21}$. Using (\ref{eq:out21BIG}) we compute $\langle \hat{a}^\dagger \hat{a} \rangle_{21}$
\begin{equation}
\label{eq:Power21}
\langle \hat{a}^\dagger \hat{a} \rangle_{21} = \frac{e^{-|\alpha|^2}|\alpha|^4}{L^2} \sum_{M=0}^{N-1} \left\lbrace \bar{r}_0^2 t_0^{2(N+M)} L + \frac{1}{8\pi} \int dk |c_M(k)|^2 + \operatorname{Re}{\left[\bar{r}_0 t_0^{N+M} c_M(0)\right]}\right\rbrace,
\end{equation}
where we note that $t_0$ and $\bar{r}_0$ are real-valued on resonance. Combining these together to fourth order in $|\alpha|$ gives
\begin{equation}
\label{eq:PowerFULLint}
\langle \hat{a}^\dagger \hat{a} \rangle = \frac{|\alpha|^2}{L} |t_0|^{2N} + \frac{|\alpha|^4}{L^2} \left\lbrace\int dx |\phi_N(x)|^2 - 2t_0^{2N} \int dx \, \phi_N(x) + \sum_{M=0}^{N-1} \frac{1}{8\pi} \int dk |c_M(k)|^2 + \operatorname{Re}{[r_0 t_0^{N+M}c_M(0)]}\right\rbrace,
\end{equation}
where we have used $t_0^{4N} - t_0^{2N} + \sum_{M=0}^{N-1} \bar{r}_0^2 t_0^{2(N+M)}=0$ and thus all the non-physical terms vanish. We note that the input power is $P_{\rm in} = |\alpha|^2/L$ and has units of photons per length. Since we have set $v_g=1$ throughout it also has units of photons per time. We perform the integrals in (\ref{eq:PowerFULLint}) numerically to produce the plot in Fig.~\ref{fig:power}(b).

\section{Asymptotics}

Here we compute the correlation function and the output power in the limit of the dynamics being dominated by detuned Fourier components. From Fig.~\ref{fig:power}(a) of the main text we have observed that when the optical depth becomes large the Fourier spectrum of the two-photon wavefunction is dominated by detuned Fourier components. By expanding to second order in $\Gamma/\Delta$ we obtain asymptotic expressions for $|\textrm{out}\rangle_{2}$, $|\textrm{out}\rangle_{21}$ and $\langle \hat{a}^\dagger \hat{a} \rangle$. Starting from Eq.~(\ref{eq:twoPhotOut}), we assume a resonant drive and ignore terms that decrease exponentially in $N$ and thus write
\begin{equation}
|\textrm{out} \rangle_{2} \sim -A \int  \frac{d \Delta \, t_{\Delta}^N t_{- \Delta}^N}{\Delta \sqrt{1+ 4\frac{\Delta^2}{\Gamma^2}}} \left\lbrace\frac{1}{\sqrt{2}} \int dx_1 dx_2 \hat{a}^\dagger (x_1) \hat{a}^\dagger (x_2) | 0 \rangle \frac{1}{\sqrt{4\Delta^2 + \Gamma^2}} \frac{\sqrt{2}}{2\pi} \left[ 2 \Delta \cos{(\Delta x)} - \Gamma \operatorname{sgn}{(x)} \sin{(\Delta x)}\right] \right\rbrace.
\end{equation}
Expanding to second order, we write $t_{\Delta}^N t_{-\Delta}^N \sim \exp[-\xi_N^2 \frac{\Gamma_{\rm tot}^2}{\Delta^2}]$, with $\xi_N=\sqrt{N \beta (1-\beta)}$ and thus our assumption of detuned frequencies dominating implies $N \beta(1-\beta) = \xi_N^2 \gg 1$. We thus have
\begin{equation}
\label{eq:out2Asymp}
|\textrm{out} \rangle_{2} \sim - A\frac{\Gamma}{4\pi} \int dx_1 dx_2 \hat{a}^\dagger (x_1) \hat{a}^\dagger (x_2) | 0 \rangle  \int d \Delta \frac{e^{-\xi^2_N \Gamma_{\rm tot}^2/\Delta^2}}{\Delta^2} \cos{(\Delta x)},
\end{equation}
and we define
\begin{equation}
F_N(x) =  \int d \Delta \frac{e^{-\xi^2_N \Gamma_{\rm tot}^2/\Delta^2}}{\Delta^2} \cos{(\Delta x)} = \frac{\sqrt{\pi}}{\Gamma_{\rm tot} \xi_N} {}_0F_2\left({\frac{1}{2}, \frac{1}{2}};{\frac{\xi_N^2 \Gamma_{\rm tot}^2 x^2}{4}}\right) - \pi |x|\, {}_0F_2\left(1, \frac{3}{2};{\frac{\xi_N^2 \Gamma_{\rm tot}^2 x^2}{4}}\right),
\end{equation}
where ${}_0 F_2$ is the generalized hypergeometric function. From this we obtain $g^{(2)}(x) \sim \frac{\Gamma^2}{4\pi^2} \frac{|F_N(x)|^2}{(1-2\beta)^{4N}}$. We furthermore obtain a $k$-space representation of (\ref{eq:out2Asymp}) as
\begin{equation}
\label{eq:out2AsympKSPACE}
|\textrm{out} \rangle_{2} \sim - A\frac{\Gamma}{2} \int dk_1 dk_2 \hat{a}^\dagger (k_1) \hat{a}^\dagger (k_2) | 0 \rangle  \delta(E_k) \frac{e^{-\xi^2_N \Gamma_{\rm tot}^2/\Delta_k^2}}{\Delta_k^2},
\end{equation}
where we define $\phi_N^{\textrm{asymp}}(\Delta_k) = \Gamma/\Delta_k^2 \exp{[-\xi_N^2 \Gamma_{\rm tot}^2 /\Delta_k^2]}$. With the two-photon wavefunction at hand we easily obtain the two-photon contribution to the nonlinear part of the power
\begin{equation}
\label{eq:asympPower2}
\langle \hat{a}^{\dagger} \hat{a} \rangle_{2} \sim  \frac{P_{\rm in}}{P_{\rm sat}} \frac{1}{8\sqrt{2\pi}} \frac{\beta}{\xi_N^{3}}.
\end{equation}

We now compute the output power $\langle \hat{a}^\dagger \hat{a} \rangle_{21}$ using (\ref{eq:Power21}) and $\phi_N^{\textrm{asymp}}(\Delta_k)$. We do this by first obtaining an asymptotic expression for $| \textrm{out} \rangle_{21}$. We consider terms scaling sub-exponentially in $N$, and thus drop the terms that are exponential in $N$. This leaves the contribution which is proportional to $c_M(k)$ (see (\ref{eq:c_M})). Within the asymptotic limit, the contribution here from $\Phi_M$ is smaller than the other terms. This term contains processes where one photon is scattered out of the waveguide while the other is transmitted through a correlated nonlinear process. This is unlikely to occur because the correlated wavefunction $\phi_N^{\textrm{asymp}}(\Delta_k)$ is dominated by detuned Fourier components and is unlikely to interact nonlinearly. We are thus left with
\begin{equation}
\label{eq:out21Asymp}
|\textrm{out}\rangle_{21} = \frac A 2  \sum_{M=0}^{N-1} \int dk \, \hat{a}_R^\dagger (k) \hat{a}_L^{\dagger (M+1)} (-k) | 0 \rangle \, t_{k}^{N-M-1}\left\lbrace - 2 \bar{r}_{-k}t_{k} \phi_M^{\textrm{asymp}}(k) + \frac{i\beta\sqrt{\Gamma (1-\beta)}}{\pi} \bar{s}_{k} \bar{s}_{-k} 4 \pi \bar{s}_0 t_0^{2M} \right\rbrace.
\end{equation}
The first term contains the asymptotic form $\phi_M^{\textrm{asymp}}(k)$ interacting linearly with emitters and the second term quantifies the correlated interactions, i.e. the two photons interact in an uncorrelated manner for the first $M$ emitters and then interact in a correlated manner on the $M+1$th emitter through the $\hat{S}_{12}$ scattering term. Computing $\langle \hat{a}^\dagger \hat{a} \rangle_{21}$ leads to three integrals: the modulus square of the first and second terms in (\ref{eq:out21Asymp}) and the cross term. We have found that the cross term does not contribute to leading order and we thus focus on the other two. First, we have
\begin{equation}
\begin{split}
\int dk |r_{-k}|^2 |t_k|^{2(N-M)} |\phi_M^{\textrm{asymp}}(k)|^2 &\sim \beta (1-\beta)\Gamma_{\rm tot}^2 \Gamma^2 \int dk \frac{e^{-\xi_{N+M}^2 \Gamma_{\rm tot}^2/k^2}}{k^6}\\
&= \frac{\beta^3(1-\beta) 3 \sqrt{\pi}}{ \xi_{N+M}^5 \Gamma_{\rm tot} },
\end{split}
\end{equation}
where we have used $|t_k|^{2N} \sim \exp{[-\xi^2_N \frac{\Gamma_{\rm tot}^2}{k^2}]}$, and the integral $\int dk e^{-c^2/k^2}/k^{2n} = \Gamma(n-\frac{1}{2})/c^{2n-1}$, where $\Gamma(n)$ is the Gamma function. The contribution from the modulus square of the second term in (\ref{eq:out21Asymp}) decays exponentially with $M$ and is thus dominated by terms $M \ll N$. This allows writing to leading order
\begin{equation}
\begin{split}
\int dk |\bar{s}_k \bar{s}_{-k}|^2 |t_k|^{2(N-M-1)} &\sim \int dk \, e^{-\xi_{N-M-1}^2 \Gamma_{\rm tot}^2 /k^2} \frac{\Gamma_{\rm tot}^2}{k^4}\\
&\sim \frac{\sqrt{\pi}}{2 \, \xi_N^3 \Gamma_{\rm tot} }.
\end{split}
\end{equation}
Using these integrals and (\ref{eq:out21Asymp}) we get
\begin{equation}
\begin{split}
\label{eq:asympPower21}
\frac{\langle \hat{a}^\dagger \hat{a} \rangle_{21}}{P_{\rm in}} &\sim \frac{P_{\rm in}}{8 \pi} \sum_{M=0}^{N-1} \left\lbrace \frac{\beta^3(1-\beta) 3 \sqrt{\pi}}{ \xi_{N+M}^5 \Gamma_{\rm tot} } + 64 \beta^3 (1-\beta) t_0^{4M} \frac{\sqrt{\pi}}{2 \, \xi_N^3 \Gamma_{\rm tot} }\right\rbrace\\
&\sim \frac{P_{\rm in}}{P_{\rm sat}}\left[ \frac{2\sqrt{2}-1}{8 \sqrt{2 \pi}} + \frac{1}{2\sqrt{\pi}} \frac{1}{1-2\beta(1-\beta)} \right] \frac{\beta}{\xi_N^3}
\end{split}
\end{equation}
where, for the first term, we have used $\sum_{M=0}^{N-1}1/(N+M)^{5/2} = \operatorname{\zeta}{(\frac 5 2, N)} - \operatorname{\zeta}{(\frac 5 2, 2N)} \sim \left(\frac{2}{3} - \frac{1}{3\sqrt{2}} \right) \frac{1}{N^{3/2}}$ for $N \gg 1$, where $\operatorname{\zeta}{(s, a)}$ is the Hurwitz zeta function, and for the second term we extended the summation to $\infty$ and computed the geometric series. Combining (\ref{eq:asympPower2}), (\ref{eq:asympPower21}), and the linear contribution gives Eq.~\ref{eq:powerAsymp} in the main text.

\section{Computing the Integrals}

In this section we compute the values of the integrals used throughout the manuscript. The first integral is
\begin{equation}
\label{eq:I1}
\begin{split}
I_1 = \int d \Delta \frac{t_{k_0+\Delta}^N t_{k_0-\Delta}^N}{\Gamma^2 + 4 \Delta^2} \cos{(\Delta x)} = \frac 1 2\int d \Delta \frac{e^{i \Delta |x|} + e^{- i \Delta |x|}}{\Gamma^2 + 4 \Delta^2} \left(\frac{\Delta + k_0 + i \Gamma (1-2\beta)/2\beta}{\Delta + k_0 + i \Gamma/2\beta}\right)^N \left(\frac{\Delta - k_0 - i \Gamma (1-2\beta)/2\beta}{\Delta - k_0 - i \Gamma/2\beta}\right)^N.
\end{split}
\end{equation}
By extending the integrand over the entire complex plane and using a contour that is closed in the upper or lower half of the complex plane the above integral can be computing using the Residue Theorem. This gives
\begin{equation}
\begin{split}
I_1 = \frac{\pi}{4 \Gamma}&\left\{2 \tilde{t}_{2k_0}^N e^{-\frac{\Gamma}{2}|x|}  + \frac{i\Gamma}{(N-1)!} \frac{d^{N-1}}{dz^{N-1}} \left[ \frac{t_{k_0 + z}^N (z-k_0 - i\Gamma (1-2\beta)/2\beta)^N e^{i z |x|}}{z^2 + \Gamma^2/4}\right]_{z=k_0 + \frac{i \Gamma}{2 \beta}} \right. \\
& \left. - \frac{i\Gamma}{(N-1)!} \frac{d^{N-1}}{dz^{N-1}} \left[ \frac{t_{k_0 - z}^N (z+k_0 +\Gamma (1-2\beta)/2\beta)^N e^{-i z |x|}}{z^2 + \Gamma^2/4}\right]_{z=-k_0 - \frac{i \Gamma}{2 \beta}} \right\},
\end{split}
\end{equation}
where we have used $t_{k_0 + \frac{i\Gamma}{2}} t_{k_0 - \frac{i\Gamma}{2}} = \tilde{t}_{2k_0}$.

The second integral we compute is
\begin{equation}
\label{eq:I2}
\begin{split}
I_2 = \operatorname{sgn}{(x)}\int \frac{d \Delta}{\Delta} \frac{t_{k_0+\Delta}^N t_{k_0-\Delta}^N}{\Gamma^2 + 4 \Delta^2} \sin{(\Delta x)} = \frac{1}{2i} &\int \frac{d \Delta}{\Delta} \frac{e^{i \Delta |x|} - e^{- i \Delta |x|}}{\Gamma^2 + 4 \Delta^2} \left(\frac{\Delta + k_0 + i \Gamma (1-2\beta)/2\beta}{\Delta + k_0 + i \Gamma/2\beta}\right)^N\\
&\times \left(\frac{\Delta - k_0 - i \Gamma (1-2\beta)/2\beta}{\Delta - k_0 - i \Gamma/2\beta}\right)^N.
\end{split}
\end{equation}
Using the same approach as for $I_1$ we obtain
\begin{equation}
\begin{split}
I_2 = \frac{i \pi}{4 \Gamma}&\left\{-\frac{4i}{\Gamma}t_{k_0}^{2N} + \frac{4i}{\Gamma} \tilde{t}_{2k_0}^N e^{-\frac{\Gamma}{2}|x|}  - \frac{i\Gamma}{(N-1)!} \frac{d^{N-1}}{dz^{N-1}} \left[ \frac{t_{k_0 + z}^N (z-k_0 - i\Gamma (1-2\beta)/2\beta)^N e^{i z |x|}}{z(z^2 + \Gamma^2/4)}\right]_{z=k_0 + \frac{i \Gamma}{2 \beta}} \right. \\
& \left. - \frac{i\Gamma}{(N-1)!} \frac{d^{N-1}}{dz^{N-1}} \left[ \frac{t_{k_0 - z}^N (z+k_0 +\Gamma (1-2\beta)/2\beta)^N e^{-i z |x|}}{z(z^2 + \Gamma^2/4)}\right]_{z=-k_0 - \frac{i \Gamma}{2 \beta}} \right\}.
\end{split}
\end{equation}

The third integral, which we compute for a resonant drive, is
\begin{equation}
\label{eq:I3}
\begin{split}
\Phi_N &= \int dk (\bar{s}_k + \bar{s}_{-k})\phi_M(k) = \frac{1}{(N-1)!}\sum_{j=0}^{N-1} \binom{N-1}{j} F_0(N,j) \int dk (\bar{s}_k + \bar{s}_{-k}) \chi_{N-1-j} (k)\\
&= \Gamma \sum_{j=0}^{N-1} \frac{ F_0(N,j)}{j!} \int dk (\bar{s}_k + \bar{s}_{-k}) \left[\frac{(k - \frac{i\Gamma}{2 \beta})^{j-N}}{k(k-\frac{i\Gamma}{2})} + \frac{(-k - \frac{i\Gamma}{2 \beta})^{j-N}}{k(k+\frac{i\Gamma}{2})} - \frac{2}{k^2 + \frac{\Gamma^2}{4}} \left(\frac{2i\beta}{\Gamma(1-\beta)} \right)^{N-j}\right]
\end{split}
\end{equation}
The third term in the square brackets is evaluated easily
\begin{equation}
\Phi_N^{(3)} =  \int dk (\bar{s}_k + \bar{s}_{-k}) \frac{2 \Gamma}{k^2 + \frac{\Gamma^2}{4}} = \frac{-16i\pi \sqrt{\beta/\Gamma}}{1+\beta}.
\end{equation}
The first two terms in the square brackets of (\ref{eq:I3}) can be combined to give an integral of the form
\begin{equation}
\begin{split}
\Phi_N^{(1)} +\Phi_N^{(2)} =  \int dk \frac{\Gamma (\bar{s}_k + \bar{s}_{-k})}{k (k^2+\frac{\Gamma^2}{4})(k^2+\frac{\Gamma^2}{4\beta^2})^{N-j}} \left\lbrace k \left[ \left(k+\frac{i \Gamma}{2\beta}\right)^{N-j}+(-1)^{N-j}\left(k-\frac{i\Gamma}{2\beta}\right)^{N-j} \right. \right] \\
+\left. \frac{i\Gamma}{2}\left[ \left(k+\frac{i \Gamma}{2\beta}\right)^{N-j}-(-1)^{N-j}\left(k-\frac{i\Gamma}{2\beta}\right)^{N-j}\right]\right\rbrace.
\end{split}
\end{equation}
Using a binomial series to expand the powers in the bracers, the entire expression can be written as
\begin{equation}
\begin{split}
\Phi_N^{(1)} +\Phi_N^{(2)} =  \sum_{m=0}^{N-j}\binom{N-j}{m}\left( \frac{i\Gamma}{2\beta}\right)^{N-j-m} I_{m,N-j},
\end{split}
\end{equation}
where
\begin{equation}
I_{m,N-j} = \begin{cases}
\int dk \frac{2 \Gamma (\bar{s}_k + \bar{s}_{-k}) k^m}{\left(k^2 + \frac{\Gamma^2}{4} \right)\left(k^2 + \frac{\Gamma^2}{4\beta} \right)^{N-j}} &\mbox{if } m \in \textrm{even} \\
\int dk \frac{i \Gamma^2 (\bar{s}_k + \bar{s}_{-k}) k^{m-1}}{\left(k^2 + \frac{\Gamma^2}{4} \right)\left(k^2 + \frac{\Gamma^2}{4\beta} \right)^{N-j}}  &\mbox{if } m \in \textrm{odd},
\end{cases}
\end{equation}
which can be evaluated to give
\begin{equation}
I_{k,j} = \begin{cases}
64 \pi (-1)^{j+1} i^{k+1} 2^{2j-k} \beta^{\frac{3}{2}+2j} \Gamma^{-\frac{5}{2}-2j} \left[ \beta(-1+\beta^2)^{-2-j}\Gamma^k - \Gamma_{\rm tot}^k \frac{\operatorname{\Gamma}{\left(\frac{k+1}{2} \right)}}{\operatorname{\Gamma}{\left(j+2 \right)}}  \pFtildeq{2}{1}{1,\frac{1+k}{2}}{\frac{1}{2}(-1-2j+k)}{\frac{1}{\beta^2}} \right] &\mbox{if } k \in \textrm{even} \\
64 \Gamma_{\rm tot}^{-\frac 5 2 - 2j} 2^{2j-k} \left[\Gamma_{\rm tot}^k \operatorname{\Gamma}{\left(2+j-\frac k 2 \right)} \operatorname{\Gamma}{\left( \frac k 2 \right)} \pFq{2}{1}{1,\frac{k}{2}}{\frac{1}{2}(-2-2j+k)}{\frac{1}{\beta^2}} + (-1)^{\frac{2j+k-1}{2}} \pi (-1+\beta^2)^{-2-j} \Gamma^k \right]  &\mbox{if } k \in \textrm{odd},
\end{cases}
\end{equation}
where ${}_2 F_{1}$ is Gauss's Hypergeometric function and ${}_2 \tilde{F}_{1}$ is Gauss's regularized Hypergeometric function. Putting everything together we get
\begin{equation}
\begin{split}
\Phi_N = \sum_{j=0}^{N-1} \frac{F_0(N,j)}{j!} \left[ \frac{16i \pi \sqrt{\beta/\Gamma}}{1+\beta} \left(\frac{2 i \beta}{\Gamma (1-\beta)} \right)^{N-j} + \sum_{m=0}^{N-j} \binom{N-j}{m} \left(\frac{i \Gamma}{2 \beta} \right)^{N-j-m} I_{m,N-j} \right]
\end{split}
\end{equation}

\section{Computing the Differentials}

In this section we compute the differentials that emerge from the evaluation of the residue of the $N$th order poles in integrals $I_1$ and $I_2$. In total there are four differentials in (\ref{eq:psiDiffer}). Here we detail the steps we use to evaluate these in terms of generalized functions.

In this section we make extensive use of the general Leibniz rule of differentiation
\begin{equation}
\frac{d^n}{d x^n}\left[ f(x) g(x) \right]_{x=x_0} = \sum_{i=0}^n \binom{n}{i} \left. \frac{d^i f(x)}{d x^i}  \right|_{x=x_0} \left. \frac{d^{n-i} g(x)}{d x^{n-i}} \right|_{x=x_0}.
\end{equation}
We start with the first differential in (\ref{eq:psiDiffer}) and write it in a compact form
\begin{equation}
\frac{1}{(N-1)!}\frac{d^{N-1}}{d z^{N-1}} \left[t_{k_0+z}^N \left(z-k_0- \frac{i \Gamma}{2\beta} (1-2\beta)\right)^N \frac{e^{i z |x|}}{z^2 + \Gamma^2/4} \right]_{z=k_0+\frac{i \Gamma}{2\beta}} \equiv \frac{1}{(N-1)!}\frac{d^{N-1}}{d z^{N-1}} \left[\left[f(z)\right]^N g(z,x) \right]_{z=\gamma},
\end{equation}
where $\gamma=k_0 + i\Gamma/2\beta$, $f(z) = t_{k_0+z} (z-k_0- \frac{i \Gamma}{2\beta} (1-2\beta))$, and $g(z,x)=e^{i z |x|}/(z^2 + \Gamma^2/4)$. Using Leibniz's rule this becomes
\begin{equation}
\frac{1}{(N-1)!}\frac{d^{N-1}}{d z^{N-1}} \left[\left[f(z)\right]^N g(z,x) \right]_{z=\gamma} = \frac{1}{(N-1)!}\sum_{m=0}^{N-1} \binom{N-1}{m} \left. \frac{d^m \left[f(z)\right]^N}{d z^m}  \right|_{z=\gamma} \left. \frac{d^{N-1-m} g(z,x)}{d z^{N-1-m}} \right|_{z=\gamma}
\end{equation}
We start by evaluating
\begin{equation}
F_{k_0}(N,m) = \left. \frac{d^m \left[f(z)\right]^N}{d z^m}  \right|_{z=\gamma} = \left. \frac{d^m}{d z^m}\left[ \frac{z^2 - a^2}{z + \gamma}\right]^N \right|_{z=\gamma},
\end{equation}
where $a=k_0 + i\Gamma(1-2\beta)/2\beta$. In order to compute this we first compute
\begin{equation}
\label{eq:nthDerofFirstPower}
\left. \frac{d^n}{d z^n} \left[ \frac{z^2 - a^2}{z + \gamma}\right] \right|_{z=\gamma} = (-1)^n n! \, (2\gamma)^{-n} \left( \frac{\gamma^2-a^2}{2\gamma} - 2\gamma \, \delta_{n-1}  \right),
\end{equation}
where $\delta_i$ is the Kronecker delta and takes values $\delta_i=0$ for $i\neq0$ and $\delta_i=1$ for $i=0$. We can now use Leibniz's rule recursively to express
\begin{equation}
\begin{split}
\label{eq:nthDerFormula}
F_{k_0}(N,m)& = \sum_{i_1=0}^m \sum_{i_2=0}^{m-i_1} \ldots \sum_{i_{N-1}=0}^{m-i_1-i_2-\ldots - i_{N-2}} \binom{m}{i_1} \binom{m-i_1}{i_2} \ldots \binom{m-i_1-i_2-\ldots - i_{N-2}}{i_{N-1}}\\
&\times  f^{(m-i_1-i_2-\ldots-i_{N-1})}(\gamma) \prod_{j=1}^{N-1}f^{(i_j)}(\gamma),
\end{split}
\end{equation}
where $f^{(n)}(x_0)$ is the $n$th derivative of $f$ evaluated at $x_0$. We now substitute (\ref{eq:nthDerofFirstPower}) into (\ref{eq:nthDerFormula}), which, after some manipulation gives
\begin{equation}
\begin{split}
\label{eq:bigExpF}
F_{k_0}(N,m )& = (-1)^m (2\gamma)^{-m} m! \sum_{i_1}^m \sum_{i_2=0}^{m-i_1} \ldots \sum_{i_{N-1}=0}^{m-i_1-i_2 \ldots i_{N-2}} \left[ \frac{\gamma^2-a^2}{2\gamma} - 2\gamma \, \delta_{i_1-1}  \right] \left[ \frac{\gamma^2-a^2}{2\gamma} - 2\gamma \, \delta_{i_2-1}  \right]\\
&\times \ldots \times \left[ \frac{\gamma^2-a^2}{2\gamma} - 2\gamma \, \delta_{i_N-1}  \right]\left[ \frac{\gamma^2-a^2}{2\gamma} - 2\gamma \, \delta_{m-i_1-i_2-\ldots-i_{N-1}-1}  \right]\\
&= (-1)^m (2\gamma)^{-m} m! \sum_{i_1}^m \sum_{i_2=0}^{m-i_1} \ldots \sum_{i_{N-1}=0}^{m-i_1-i_2 \ldots i_{N-2}} \frac{\gamma^2-a^2}{2\gamma} \left[ \frac{\gamma^2-a^2}{2\gamma} - 2\gamma \, \delta_{i_1-1}  \right] \left[ \frac{\gamma^2-a^2}{2\gamma} - 2\gamma \, \delta_{i_2-1}  \right]\\
&\times \ldots \times \left[ \frac{\gamma^2-a^2}{2\gamma} - 2\gamma \, \delta_{i_N-1}  \right]\\
&+ (-1)^m (2\gamma)^{-m} m! \sum_{i_1}^m \sum_{i_2=0}^{m-i_1} \ldots \sum_{i_{N-1}=0}^{m-i_1-i_2 \ldots i_{N-2}} - 2\gamma \, \delta_{m-i_1-i_2-\ldots-i_{N-1}-1} \left[ \frac{\gamma^2-a^2}{2\gamma} - 2\gamma \, \delta_{i_1-1}  \right] \left[ \frac{\gamma^2-a^2}{2\gamma} - 2\gamma \, \delta_{i_2-1}  \right]\\
&\times \ldots \times \left[ \frac{\gamma^2-a^2}{2\gamma} - 2\gamma \, \delta_{i_N-1}  \right],
\end{split}
\end{equation}
where in the last equality we have split the expression into two terms. These two terms can be written compactly as polynomials using the binomial theorem. We write the first as
\begin{equation}
\frac{\gamma^2 - a^2}{2\gamma}\sum_{k=0}^{N-1} \binom{N-1}{k}\left( \frac{\gamma^2-a^2}{2\gamma}\right)^{N-1-k} (-1)^k (2\gamma)^k V(N,m,k),
\end{equation}
and
\begin{equation}
V(N,m,k) = \sum_{i_1=0}^m \sum_{i_2=0}^{m-i_1} \ldots \sum_{i_{N-1}=0}^{m-i_1-i_2-\ldots -i_{N-1}} \delta_{i_{x_1}-1} \delta_{i_{x_2}-1} \ldots \delta_{i_{x_k}-1},
\end{equation}
where there are $k$ $\delta$ factors, and the subscript $x_j$ refers to any of the $\delta$ terms when writing the above as a polynomial. Importantly, we can write the form using a binomial expansion only because the value of $V(N,m,k)$ is independent of the subscript of the $\delta$ functions and only depends on the total number of them $k$. We have found that
\begin{equation}
V(N,m,k) = \theta(m-k)\frac{(N-1+m-2k)!}{(m-k)! (N-1-k)!},
\end{equation}
where $\theta(j)$ is the unit step function where $\theta(j)=1$ for $j \geq 0$ and $\theta(j)=0$ otherwise, and $m \leq N-1$ and $k \leq N-1$. We can similarly write the second term of (\ref{eq:bigExpF}) as
\begin{equation}
-2 \gamma \sum_{k=0}^{N-1} \binom{N-1}{k}\left(\frac{\gamma^2 - a^2}{2\gamma}\right)^{N-1-k}(-1)^k (2\gamma)^k D(N,m,k),
\end{equation}
where
\begin{equation}
\begin{split}
D(N,m,k) &= \sum_{i_1=0}^m \sum_{i_2=0}^{m-i_1} \ldots \sum_{i_{N-1}=0}^{m-i_1-i_2-\ldots -i_{N-1}} \delta_{m-i_1 - i_2 - \ldots i_{N-1}-1}\delta_{i_{x_1}-1} \, \delta_{i_{x_2}-1} \ldots \delta_{i_{x_k}-1}\\
&= \theta(m-k-1) \frac{(m+N-3-2k)!}{(m-1-k)! (N-k-2)!},
\end{split}
\end{equation}
and again we can use this form because the value of the sum only depends on the number of $\delta$ factors $k$. Using (\ref{eq:bigExpF}) and combining the two terms together we get
\begin{equation}
\begin{split}
\label{eq:3F2}
F_{k_0}(N,m) &= (-1)^m (2\gamma)^{-m} m! \sum_{k=0}^m \frac{N}{N-k}\binom{N-1}{k}\left(\frac{\gamma^2 - a^2}{2\gamma}\right)^{N-k}(-1)^k(2\gamma)^k \frac{(N-1+m-2k)!}{(N-k-1)! (m-k)!}\\
&=(-1)^m (2\gamma)^{-m}\left(\frac{\gamma^2-a^2}{2\gamma}\right)^N \frac{(N+m-1)!}{(N-1)!} \pFq{3}{2}{-m,1-N,-N}{\frac{1-m-N}{2},\frac{2-m-N}{2}}{\frac{\gamma^2}{\gamma^2-a^2}},
\end{split}
\end{equation}
where ${}_3F_2$ is the generalized Hypergeometric function.

We now move to computing the derivative
\begin{equation}
\begin{split}
\frac{d^m}{dz^m}\left[\frac{e^{i z |x|}}{z^2 + \Gamma^2/4}\right]_{z=\gamma} &= \frac{e^{i \gamma |x|}}{i \Gamma} \sum_{j=0}^m \binom{m}{j}(i |x|)^{m-j} (-1)^j j! \left[ \left(\gamma - \frac{i\Gamma}{2} \right)^{-1-j} -  \left(\gamma + \frac{i\Gamma}{2} \right)^{-1-j}\right]\\
&=\frac{1}{i \Gamma}\left\lbrace e^{\frac \Gamma 2 |x|} \left(-\frac{i \Gamma}{2}-\gamma \right)^{-m-1} \Gamma_{m+1}\left[-i \left(\gamma + \frac{i \Gamma}{2}\right) |x|\right] \right.\\
&\left. - e^{-\frac \Gamma 2 |x|}  \left(\frac{i \Gamma}{2}-\gamma \right)^{-m-1}   \Gamma_{m+1}\left[-i \left(\gamma - \frac{i \Gamma}{2}\right) |x|\right] \right\rbrace\\
&\equiv \frac{1}{i\Gamma} \xi^{(1)}_{k_0,m}(x),
\end{split}
\end{equation}
where $\Gamma_m(x)$ is the incomplete Gamma function. We finally thus have
\begin{equation}
\begin{split}
\frac{1}{(N-1)!}\frac{d^{N-1}}{d z^{N-1}} \left[\left[f(z)\right]^N g(z,x) \right]_{z=\gamma} = \frac{1}{i\Gamma(N-1)!}\sum_{m=0}^{N-1}\binom{N-1}{m} F_{k_0}(N,m) \xi^{(1)}_{k_0,N-1-m}(x).
\end{split}
\end{equation}

The process for evaluating the remaining three differentials is almost identical to the first. The second differential in (\ref{eq:psiDiffer}) is
\begin{equation}
\frac{1}{(N-1)!}\frac{d^{N-1}}{d z^{N-1}} \left[t_{k_0-z}^N (z+k_0+ \frac{i \Gamma}{2\beta} (1-2\beta))^N \frac{e^{-i z |x|}}{z^2 + \Gamma^2/4} \right]_{z=-k_0-\frac{i \Gamma}{2\beta}} \equiv \frac{1}{(N-1)!}\frac{d^{N-1}}{d z^{N-1}} \left[\left[f_2(z)\right]^N g_2(z,x) \right]_{z=-\gamma},
\end{equation}
One can show that
\begin{equation}
\left. \frac{d^m}{dz^m}\left[f_2(z) \right]^N \right|_{z=-\gamma} = (-1)^{(N-m)} F_{k_0}(N,m),
\end{equation}
and that
\begin{equation}
\frac{d^m}{dz^m} \left[\frac{e^{-i z |x|}}{z^2+\Gamma^2/4} \right]_{z=-\gamma} = (-1)^m \frac{1}{i\Gamma} \xi^{(1)}_{k_0,m}(x),
\end{equation}
and therefore
\begin{equation}
\frac{1}{(N-1)!}\frac{d^{N-1}}{d z^{N-1}} \left[\left[f_2(z)\right]^N g_2(z,x) \right]_{z=-\gamma} = \frac{-1}{i\Gamma(N-1)!}\sum_{m=0}^{N-1}\binom{N-1}{m} F_{k_0}(N,m) \xi^{(1)}_{k_0,N-1-m}(x).
\end{equation}

The third differential has the same function raised to the $N$th power as the first, but the other factor differs, ie.,
\begin{equation}
\frac{1}{(N-1)!}\frac{d^{N-1}}{d z^{N-1}} \left[t_{k_0+z}^N (z-k_0- \frac{i \Gamma}{2\beta} (1-2\beta))^N \frac{e^{-i z |x|}}{z(z^2 + \Gamma^2/4)} \right]_{z=k_0+\frac{i \Gamma}{2\beta}} \equiv \frac{1}{(N-1)!}\frac{d^{N-1}}{d z^{N-1}} \left[\left[f(z)\right]^N g_3(z,x) \right]_{z=\gamma}.
\end{equation}
We therefore are only required to compute
\begin{equation}
\begin{split}
\frac{d^m}{dz^m} \left[\frac{e^{-i z |x|}}{z(z^2+\Gamma^2/4)} \right]_{z=\gamma} &= \frac{e^{i\gamma |x|}}{i\Gamma}\sum_{k=0}^{m} \binom{m}{k} (i|x|)^{m-k}\frac{d^k}{d z^k} \left[\frac{-1}{z(z+i\Gamma/2)}+\frac{1}{z(z-i\Gamma/2)} \right]\\
&= \frac{e^{i\gamma |x|}}{i\Gamma}\sum_{k=0}^{m} \binom{m}{k} (i|x|)^{m-k}(-1)^k k! \sum_{l=0}^k \gamma^{l-k-1} \left[\left(\gamma - \frac{i \Gamma}{2} \right)^{-1-l} - \left(\gamma + \frac{i \Gamma}{2} \right)^{-1-l} \right]\\
&= \frac{e^{i\gamma |x|}}{i\Gamma}\sum_{k=0}^{m} \binom{m}{k} (i|x|)^{m-k}(-1)^k k! \frac{\gamma^{-k-1}}{i\Gamma/2}\left[-2+ \left(\frac{\gamma - i\Gamma/2}{\gamma} \right)^{-k-1} + \left(\frac{\gamma + i\Gamma/2}{\gamma} \right)^{-k-1} \right]\\
&= \frac{1}{\Gamma^2}\left\lbrace -4 (-\gamma)^{-m-1} \Gamma_{m+1}(- i \gamma |x|) + 2\,e^{-\frac \Gamma 2 |x|} \left(-\gamma + \frac{i \Gamma}{2} \right)^{-m-1} \Gamma_{m+1}\left[-i\left( \gamma - \frac{i\Gamma}{2} \right)|x| \right] + \right.\\
&+ \left. 2\,\Gamma_{m+1}\left[-i\left( \gamma + \frac{i\Gamma}{2}\right)|x| \right] \left(-\gamma - \frac{i \Gamma}{2}\right)^{-j-1} e^{\frac \Gamma 2 |x|}\right\rbrace \equiv \frac{1}{\Gamma^2} \xi_{k_0,m}^{(3)}(x),
\end{split}
\end{equation}
and thus
\begin{equation}
\frac{1}{(N-1)!}\frac{d^{N-1}}{d z^{N-1}} \left[\left[f(z)\right]^N g_3(z,x) \right]_{z=\gamma} =  \frac{1}{\Gamma^2(N-1)!}\sum_{m=0}^{N-1}\binom{N-1}{m} F_{k_0}(N,m) \xi^{(3)}_{k_0,N-1-m}(x).
\end{equation}

Finally the fourth differential in (\ref{eq:psiDiffer}) is
\begin{equation}
\frac{1}{(N-1)!}\frac{d^{N-1}}{d z^{N-1}} \left[t_{k_0-z}^N (z+k_0+ \frac{i \Gamma}{2\beta} (1-2\beta))^N \frac{e^{-i z |x|}}{z(z^2 + \Gamma^2/4)} \right]_{z=-k_0-\frac{i \Gamma}{2\beta}} \equiv \frac{1}{(N-1)!}\frac{d^{N-1}}{d z^{N-1}} \left[\left[f_2(z)\right]^N g_4(z,x) \right]_{z=-\gamma}.
\end{equation}
One can show that
\begin{equation}
\frac{d^m}{dz^m} \left[\frac{e^{-i z |x|}}{z(z^2+\Gamma^2/4)} \right]_{z=-\gamma} = \frac{(-1)^{m+1}}{\Gamma^2} \xi^{(3)}_{k_0,m}(x),
\end{equation}
and thus
\begin{equation}
\frac{1}{(N-1)!}\frac{d^{N-1}}{d z^{N-1}} \left[\left[f_2(z)\right]^N g_4(z,x) \right]_{z=-\gamma} = \frac{1}{\Gamma^2(N-1)!}\sum_{m=0}^{N-1}\binom{N-1}{m} F_{k_0}(N,m) \xi^{(3)}_{k_0,N-1-m}(x).
\end{equation}

We can finally combine all four differentials together to get an expression for $\psi_N(x_c,x)$, which after some manipulation yields
\begin{equation}
\label{eq:psiExpression}
\psi_N(x_c,x) = e^{2i k_0 x_c}\left\lbrace t_{k_0}^{2N} - \frac{1}{(N-1)!}\sum_{n=0}^{N-1} \binom{N-1}{n} F_{k_0}(N,n) \chi_{k_0,N-1-n}(x) \right\rbrace,
\end{equation}
where
\begin{equation}
\begin{split}
\label{eq:chiExpression}
\chi_{k_0,n}(x) &= \xi_{k_0,n}^{(1)}(x)-\frac 1 2\xi_{k_0,n}^{(3)}(x)\\
&=2 (-\gamma)^{-n-1} \Gamma_{n+1}\left(-i\gamma |x| \right) - 2 e^{- \frac{\Gamma}{2} |x|} \left(-\gamma + \frac{i\Gamma}{2}\right)^{-n-1}\Gamma_{n+1}\left[-i\left(\gamma - \frac{i\Gamma}{2} \right) |x|\right],
\end{split}
\end{equation}
giving (\ref{eq:psiFull}).

\end{document}